\documentclass[aps,twocolumn,amsmath,amssymb,prd]{revtex4-1}

\usepackage{wasysym}
\usepackage{tikz}

\usepackage{color}
\usepackage{xcolor}
\usepackage{calc}
\usepackage{mathtools,graphicx}
\usepackage{pifont}
\usepackage{bm}
\usepackage{microtype}
\usepackage{booktabs}
\usepackage{times}
\usepackage[varg]{txfonts}
\usepackage[utf8]{inputenc}
\usepackage[caption=false]{subfig}
\usepackage{paralist}
\usepackage[normalem]{ulem}
\usepackage{multirow}
\usepackage{etoolbox}
\usepackage{graphics}
\allowdisplaybreaks
\DeclareFontFamily{OT1}{pzc}{}
\DeclareFontShape{OT1}{pzc}{m}{it}{<-> s * [1.10] pzcmi7t}{}
\DeclareMathAlphabet{\mathpzc}{OT1}{pzc}{m}{it}

\def\iop{i_\oplus}

\def\al{\alpha}
\def\be{\beta}
\def\ga{\gamma}
\def\de{\delta}
\def\ep{\epsilon}

\def\th{\theta}

\def\la{\lambda}

\def\rh{\rho}

\def\si{\sigma}

\def\ta{\tau}

\def\ph{\phi}

\def\ch{\chi}

\def\om{\omega}

\def\De{\Delta}

\def\Om{\Omega}
\def\mn{{\mu\nu}}

\def\fr#1#2{{{#1} \over {#2}}}

\def\frac#1#2{{\textstyle{{#1}\over {#2}}}}

\def\prt{\partial}

\def\pt#1{\phantom{#1}}

\def\sb{\overline{s}{}}

\def\stj{\sb^{\bar t \bar j}}
\def\stx{\sb^{\bar t \bar x}}
\def\sty{\sb^{\bar t \bar y}}
\def\stz{\sb^{\bar t \bar z}}

\newcommand{\beq}{\begin{equation}}
\newcommand{\eeq}{\end{equation}}
\newcommand{\bea}{\begin{eqnarray}}
\newcommand{\eea}{\end{eqnarray}}
\newcommand{\bit}{\begin{itemize}}
\newcommand{\eit}{\end{itemize}}
\newcommand{\rf}[1]{(\ref{#1})}

\begin{document}

\title{Lorentz violation and Sagnac gyroscopes}

\author{Serena Moseley,$^1$ Nicholas Scaramuzza,$^2$ Jay D.\ Tasson,$^{1}$}
\email[Corresponding author, ] {jtasson@carleton.edu}
\author{Max L.\ Trostel$^1$}

\affiliation{
$ ^1$Physics and Astronomy Department,
Carleton College,
Northfield, Minnesota 55057, USA\\
$ ^2$Physics Department,
St.\ Olaf College,
Northfield, Minnesota 55057, USA}

\date{August 2019}

\begin{abstract}

Sagnac gyroscopes with increased sensitivity are being developed
and operated with a variety of goals including the measurement of General-Relativistic effects.
We show that such systems can be used to search for Lorentz violation
within the field-theoretic framework of the Standard-Model Extension,
and that competitive sensitivities can be achieved.
Special deviations from the inverse square law of gravity
are among the phenomena that can be effectively sought with these systems.
We present the necessary equations to obtain sensitivities to Lorentz violation
in relevant experiments.

\end{abstract}

\maketitle

\section{Introduction}
\label{intro}

Sagnac interferometers \cite{sagnac1913}
have a long history as rotation sensors
and have found application in inertial guidance systems \cite{ring1}.
Increasingly,
researchers are turning to these instruments for other applications
including the measurement of geophysical effects 
and as a means of testing fundamental physics \cite{[{See, for example, }]beverini2016}.
Some such efforts are aimed at measuring General-Relativistic phenomena
including gravitomagnetic fields \cite{bosi2011}.
In General Relativity,
moving masses provide additional perturbations to spacetime
over sources at rest.
When one considers the linearized limit
of the nonlinear theory of gravity provided by General Relativity,
these effects appear as a nearly direct analogue
of the magnetic fields generated by moving charges
in electrodynamics.

Light-based Sagnac interferometers consist of counterpropagating modes
for light
in a ring-shaped interferometer.
The beat frequency between the modes
is then observed.
The noninertial frame effects
generated when the system is rotated,
as well as the effect of gravitomagnetic fields
can be understood as breaking the symmetry between the clockwise and counterclockwise
modes,
which leads to the beat signal \cite{bosi2011}.
Matter-wave Sagnac interferometers are also in use
and the analogous effect on matter waves \cite{ct}
is among the effects utilized by these devices
to sense rotation \cite{dubetsky06}.

In this work,
we demonstrate that violations of Lorentz invariance
described by the field-theoretic framework 
of the gravitational Standard-Model Extension (SME)
\cite{ck,akgrav}
can also be the source of the broken symmetry
in interferometric gyroscopes 
and can broadly mimic rotating-frame effects in such systems.
Hence sensitive interferometric gyroscopes
can also be used to search for Lorentz violation in the SME
\cite{[{For earlier discussion of the basic idea, see, }] ruggiero15,
 [{For earlier discussion of a special case, see, }] scaramuzza16}.

Lorentz invariance,
the invariance of physics
under rotations and boosts,
lies at the foundation of our current best theories:
Einstein's General Relativity
and the Standard Model of particle physics.
Hence testing Lorentz invariance to the best of our ability
is essential.
Moreover,
it is widely expected that 
General Relativity and the Standard Model,
a pair of separate theories restricted to their own domains,
are merely the low-energy limit of a single more fundamental theory
at the Planck scale.
It has been shown that Lorentz violation may arise in some candidates
for the fundamental theory \cite{ksp,*akrp1991}.
Hence a systematic search for violations of Lorentz invariance
across physics may reveal hints of the underlying theory
with present-day technology.

A comprehensive theoretical framework
is an essential tool for a systematic search.
The SME
provides that framework for Lorentz-violation searches 
\cite{ck,akgrav,[{For a review, see }]tasson14,
[{For a pedagogical introduction to Lorentz violation, see, }]bertschinger18}.
The SME is developed at the level of the action
by adding all Lorentz-violating terms to the action for known physics.
These terms consist of Lorentz-violating operators
constructed from the fields of General Relativity and the Standard Model
coupled to coefficients (or coefficient fields) for Lorentz violation.
The coefficients can then be measured or constrained
by experiment and observation.
The SME also provides a framework for theoretical study of Lorentz symmetry
\cite{qbcl2018,*akzl2019,*akbe2018,*akav2015,*akrl2001,*bonder2018,*seifert2018,*bluhm2016}

A large number of experimental and observational searches
have been performed in the context of the SME \cite{data}.
This includes considerable work in the gravity sector,
where experiments and observations have been done following
a number of phenomenological works \cite{lvpn,lvgap,kmgw,kmSR2016,*kmLG2017,*qbakrx2014,*akav2015,*mewes2019,qd5}.
Recent gravitational tests include
those found in Refs.\ \cite{qblsvel,shaol2019,cgshao2019,*gwgrb2017,*shaoGmeter2017,bourgoin2017,flowers2016}.
The tests proposed here
have the potential to compete
with the existing tests above
and complement existing discussions of Lorentz violation
in interferometric gyroscopes performed in the context
of other models and frameworks  \cite{bosi2011,ruggiero15}.

In the remainder of this work,
we demonstrate how sensitive interferometric gyroscopes
may generate additional sensitivities to gravity-sector SME coefficients.
In Sec.\ \ref{basics}
we review aspects of Lorentz violation in the SME
relevant for the development to follow.
Section \ref{ring} develops the form of the Lorentz-violation signal
in the systems of interest.
We discuss some applications of this generic result
to existing experiments and those under development in Sec.\ \ref{expt}.
Throughout this paper we use natural units except where otherwise noted
along with the other conventions of Ref.\ \cite{lvpn}.

\section{Basic Theory}
\label{basics}

The SME expansion can be thought of in analogy with a series expansion.
Terms are classified by the mass dimension $d$ of the Lorentz-violating operators
added to known physics \cite{[{For a pedagogical discussion
of mass dimension in the SME, see, }] tasson17}.
The action for the Standard Model and General Relativity
consists of dimension 3 and 4 operators.
Hence the leading Lorentz-violating corrections to known physics
are associated with operators of mass dimension 3 and 4,
which form the minimal SME.
Higher mass-dimension Lorentz-violating operators
are also of interest as models exist which generate higher mass-dimension terms
in the absence of minimal terms.
In what follows we focus primarily on the gravity sector,
for which the minimal and linearized nonminimal actions
were developed in Refs.\ \cite{akgrav} and \cite{kmgw}, respectively.

Post-Newtonian analyses have been performed
for mass-dimension 4 and 5 Lorentz-violating operators
to obtain the metric from the action
in Riemann spacetime \cite{lvpn,qd5}.
In the analysis to follow,
we consider only the leading contributions from Lorentz violation
to a post-Newtonian expansion
as the inclusion of subleading terms does not lead to additional interesting
sensitivities in the relevant experiments.
The following contributions to the metric were found
at the Newtonian level of the post-Newtonian expansion:
\bea
\nonumber
g_{00} &=& -1 + 2U + 3 \sb^{00} U 
+\sb^{jk} U^{jk}
\\
\nonumber
g_{0j} &=& -\sb^{0j}U - \sb^{0k} U^{jk} + \frac 12 {\hat Q}^j \ch
\\
g_{jk} &=& \de^{jk} 
+ (2 - \sb^{00})\de^{jk} U
\nonumber\\
&& 
+ ( \sb^{lm} \de^{jk} 
- \sb^{jl} \de^{mk}
-\sb^{kl} \de^{jm}
+ 2\sb^{00} \de^{jl} \de^{km} ) U^{lm}.
\label{sme}
\eea
Note that although these are Newtonian-order contributions to the metric,
some will appear as post-Newtonian contributions to certain observables
as they are multiplied by additional relativistic factors.
Here $U$ is the Newtonian potential,
\beq
U = G \int d^3x' \fr {\rh (\vec x', t)} {R},
\eeq
where
$G$ is Newton's constant,
$\rh$ is mass density,
and 
$R$ is the magnitude of $R^j = x^j - x'^j$,
the vector pointing from the source position $x'^j$
to the observation point $x^j$.
It is also convenient to introduce the superpotential \cite{will1993, lvpn}:
\beq
\ch = -G \int d^3x^\prime \rh ( \vec x', t) R,
\eeq
and the additional potential 
\beq
U^{jk} =\prt_j \prt_k\ch +\de_{jk} U.
\eeq
The object $\sb^\mn$ is a $d=4$ coefficient for Lorentz violation
that provides the relevant minimal gravity-sector effects.
It is symmetric and traceless,
hence minimal Lorentz violation in this limit is characterized by 9 components.
The operator ${\hat Q}^j$
is defined as
\bea
{\hat Q}^j &=& [q^{(5) 0jk0l0m}+q^{(5) n0knljm} +q^{(5) njknl0m}] \prt_k\prt_l\prt_m,
\eea
in terms of the $d=5$ coefficient for Lorentz violation $q^{(5) \mu \rh \al \nu \be \si \ga}$,
having symmetries defined in Ref.\ \cite{kmgw}.
In the analysis to follow,
the $d=5$ coefficients appear in the combinations
\bea
K_{jklm} = -\frac 16 ( q^{(5)}_{0jk0l0m} + q^{(5)}_{n0knljm} + q^{(5)}_{njknl0m} + {\rm perms} ),
\eea
where perms indicates all symmetric permutations of the indices $klm$.
We express the 15 independent combinations of $K_{jklm}$
that are observable in this work
in terms of the canonical set introduced in Ref.\ \cite{qblsvel}.
The coefficients for Lorentz violation can be understood as characterizing the amount of Lorentz violation
in the theory.
In accordance with the discussion in Ref.\ \cite{lvpn},
the coefficients for Lorentz violation satisfy
$\prt_\al \sb^\mn = 0$
and
$\prt_\de q^{(5) \mu \rh \al \nu \be \si \ga} = 0$
in the asymptotically inertial Cartesian coordinates used here.

We note in passing that the techniques presented here
can also, in principle, be used to probe Lorentz violation
in matter-gravity couplings \cite{akjt2009,lvgap}.
Lorentz-violating effects associated with the source material
can be incorporated in a straightforward way
using the post-Newtonian metric presented in Ref.\ \cite{lvgap}.
For matter-wave interferometers,
the coefficients for Lorentz violation
associated with the matter in the interferometer
are also relevant
and the associated signals
have been presented elsewhere for other interferometer configurations \cite{lvgap}.
Given the current strength of sensitivities in the matter sector,
we avoid further consideration of matter-sector Lorentz violation
in this work.

In the analysis to follow
it is convenient to introduce
three coordinate systems.
A Sun-centered celestial equatorial frame
with basis vectors $\hat Z$ aligned with Earth's spin axis,
$\hat X$ pointing toward the vernal equinox in the year 2000,
and $\hat Y$ completing the right-handed system
is the standard frame in which SME sensitivities are reported \cite{data}.
We denote the associated coordinates $T,X,Y,Z$.
We also make use of a set of coordinates
aligned with the Sun-centered coordinates and centered at the Earth
denoted $\bar t, \bar x, \bar y, \bar z$.
Finally,
we introduce a laboratory basis $\hat x, \hat y, \hat z$
in which $\hat z$ points vertically up, $\hat x$ points south,
and $\hat y$ completes the right-handed set.
We then introduce the following angles necessary to describe 
the location of any Earth-based laboratory.
Let $\th$ be the polar angle in the Earth-based coordinates
corresponding to the colatitude of the experiment,
and let $\ph$ be the corresponding azimuthal angle around the Earth
measured from the $X$-axis.
These coordinates and angles are shown in Fig.\ 1.

\begin{figure}
\includegraphics[scale=.4]{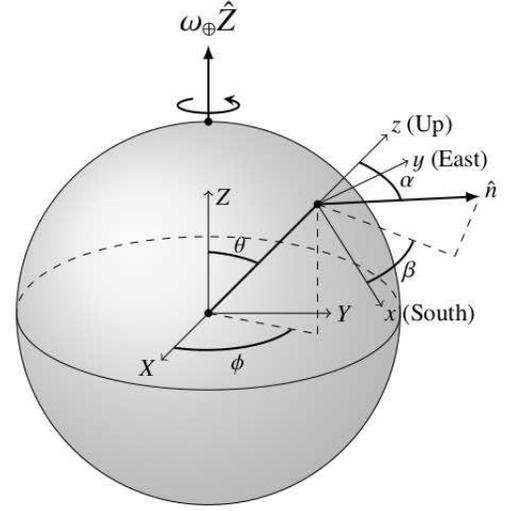}
	\caption{Diagram showing the coordinates and angles
used to describe the location and orientation of experiments
in this work. }
	\label{fig:earth}
\end{figure}

When considering experiments with Earth as the source
(approximated as spherical),
the potentials can be written
\beq
U = \fr{G M_\oplus}{\bar r},
\eeq
and
\bea
U^{\bar j \bar k} &=& 
\fr {GM_{\oplus} \bar r^{\bar j} \bar r^{\bar k} }
{\bar r^3}
-\fr {G I_{\oplus}}{3\bar r^5} 
[3 \bar r^{\bar j} \bar r^{\bar k}
-\de^{\bar j \bar k} \bar r^2],
\nonumber\\
\eea
where $M_\oplus$ is the mass of the Earth,
$\bar r^{\bar j}$
are components of a position vector
in the Earth-centered frame,
and $\bar r$ is the magnitude of position.
The quantity 
\bea
I_{\oplus} = 
\int d^3 \bar r' \rh(\vec{\bar r}^{\hskip 2 pt \prime}) \bar r'^2
\label{inertia}
\eea
is the spherical moment of inertia of the Earth.
For later convenience we define the scaled spherical moment,
\beq
i_\oplus = \fr{I_{\oplus}}{M_\oplus R_\oplus^2},
\eeq
which has a value of approximately 0.50 \cite{iop},
as well as the symbol
\beq
i_n = 1 + n i_\oplus.
\eeq
Here $R_\oplus$ is the radius of the Earth
and $n$ is a real number.

\section{Gyroscope Analysis}
\label{ring}

In this section,
we review the aspects of interferometric gyroscope measurements
of spacetime properties
that are relevant for our analysis of Lorentz violation
before applying these tools
to the general theory of Lorentz violation above.

\subsection{Spacetime probe}

\subsubsection{Light}

We begin in analogy with a metric-based approach
to the analysis of rotating-frame
and gravitomagnetic effects
in photon-based systems \cite{bosi2011}.
Lightlike trajectories satisfy the null condition 
\beq
0 = g_\mn dx^\mu dx^\nu.
\eeq
When $g_{0j}$ is nonzero
in the proper frame of the experiment,
2 solutions for the time taken for a photon
to travel around a loop emerge from the null condition
for a suitably chosen loop.
The difference in these times 
is the difference in the time taken
to go around the loop in opposite directions. 
Considering measurements made in a laboratory at rest
with coordinates $x_L^j$ in a stationary metric,
the proper time difference measured in this frame can be written
\beq
\De \tau = 2 \sqrt{ g_{00}(x_L^j)} \oint \fr{g_{0j}}{g_{00}} dx^j.
\label{det}
\eeq
However in the analysis to follow,
we consider the leading effects involving the first power of coefficients for Lorentz violation.
Higher powers of Lorentz violation as well as
Lorentz violation suppressed by post-Newtonian effects beyond the Newtonian level
or noninertial frame effects associated with the rotation of an Earth-based laboratory
are significantly smaller relative to the leading effects of Lorentz violation,
and they are not considered.
Further,
we do not present the standard Lorentz-invariant contributions
to the interferometric-gyroscope signal as these have been well studied elsewhere \cite{bosi2011}.
With these specializations
and the metric \rf{sme},
it suffices to write
\beq
\De \tau \approx 2 \oint g_{0j} dx^j.
\label{det}
\eeq
Continuing at leading order, 
the phase difference in the counterpropagating modes
per orbit
is 
\beq
\De \psi = 2 \pi \fr{\De \tau}{\la},
\label{phase}
\eeq
where $\la$ is the wavelength of the light.
Hence the number of orbits per beat cycle
is 
\beq
N = \fr{\la}{\De \tau}.
\eeq
The beat period can be written as
\beq
T = N P,
\eeq
where $P$ is the perimeter of the loop.
Thus the beat frequency is 
\beq
f_b = \fr{\De \tau}{\la P}.
\eeq

For the cases of interest here,
the integral in Eq.\ \rf{det} is most straightforwardly evaluated
by analogy with Ampere's law,
with $g_{0j}$ playing the role of the magnetic field.
The curl of $g_{0j}$ then plays the role of the current density,
which we call $2 \vec \Omega$
due to its relation to the angular velocity of the lab
in the context of Sagnac experiments.
Denoting with $\vec \Om^{(s)}$ the rotating frame contribution to $\vec \Om$,
one finds, for example,
\beq
\vec \Om^{(s)} =\vec \om_\oplus
\eeq
for a device at rest on the Earth,
where $\vec \om_\oplus$ is Earth's angular velocity.
For other contributions to $g_{0j}$,
$\vec \Om$ can be understood as an effective rotation rate,
an analogy that is useful in estimating sensitivities
to fundamental-physics effects.

Continuing by analogy with Ampere's law,
$\De \tau$ can be written as the integral
over the area enclosed by the loop as follows
\beq
\De \tau = 4 \int \vec \Omega \cdot \hat n dA,
\label{omtau}
\eeq
where $\hat n$ is a unit vector normal to the loop.
To evaluate $\De \tau$ 
it is convenient to introduce several angles.
Let $\al$ be a polar angle in the laboratory measured from the laboratory $z$-axis
to $\hat n$,
and let $\be$ be an azimuthal angle around the laboratory vertical
from the laboratory $x$-axis to $\hat n$.
Figure 1 shows these definitions.
For the cases of interest,
the area of the interferometer is sufficiently small
that $\vec \Om$ can be taken as uniform over its extent
and the integral in Eq.\ \rf{omtau}
can be evaluated as a simple product.

\subsubsection{Matter waves}

Though we consider light-based Sagnac gyroscopes in detail above,
the notion of $\vec \Om$ as an effective rotation rate
typically applies to matter-wave Sagnac gyroscopes as well.
In atom-interferometer gyroscopes,
an atom beam is typically split and recombined using light pulses
in such a way that the beam paths enclose an area \cite{riehle91,*gustavson97,*stockton11,*savoie18}.
In these systems,
the phase difference of the beams at recombination
provides the signal.

In the context of Sagnac-like signals
at leading order in $\vec \Om$,
there are 2 relevant mechanisms by which the phase is impacted:
the phase accumulated during free propagation between light pulses
and the effect of the light pulses \cite{ct}.
The free propagation contribution
can be calculated by integrating the Lagrangian around the loop.
This calculation is completely analogous
to that done for photons above,
and it leads to a phase difference of the form of Eq.\ \rf{phase}
with the wavelength given by the Compton wavelength
of the interfering particles of mass $m$:
\beq
\la \rightarrow \fr{1}{m}, 
\eeq
and $\De \ta$ given by Eq.\ \rf{omtau}.
This result
is sometimes conveniently expressed \cite{dubetsky06} in terms of the total
time the particle spends in the interferometer $T$,
the effective wave vector of the pulses $\vec k$,
and the initial momentum of the atoms $\vec p$,
in which case it is proportional to 
$\fr{1}{m} (\vec \Om \times \vec k) \cdot \vec p T^2$.

The phase imprint due to the light pulses
depends on the locations
at which the light-atom interactions occur.
Since $\vec \Om$ alters the path of the atoms through the equation of motion,
a signal also arises here.
For a particle in an Earth-based laboratory,
the relevant parts of the equation of motion are
\beq
\vec a = \vec g - 2 \vec \Om \times \vec v,
\eeq
where $\vec v$ is the velocity of the atoms and $\vec g$
is the local gravitational field.
Applying the solutions to this equation
to atom interferometers \cite{dubetsky06}
yields a leading phase shift proportional to $\vec k \cdot (\vec g \times \vec \Om) T^3$.
Though typically smaller than the free-propagation signal,
the interaction signal offers other advantages
and is sometimes used as the dominant rotation-sensing
effect.

\subsection{Lorentz violation}

In this subsection,
we apply the general gyroscope results above
to Lorentz violation in the SME.
We first consider the minimal SME,
and then higher mass-dimension terms.

\begin{table*}
\label{tab:sme_coeff}
\begin{tabular}{|l|c|c|c|}
  \hline
  Coeff.	&$\Om^{(5)}_r$	 	&$\Om^{(5)}_\th$	 	&	 $\Om^{(5)}_\ph$			\\
  \hline
$K_{XXXY}$ &	$-12 (3 i_{-5/3} c3\th  + 5 i_{-3/5} c\th  - 12 i_{-5/3} c2\ph c\th  s^2\th)$&	$-3 (6 i_{-5} c^2\ph s3\th  + 7 i_{-3/7} s\th  + 15 i_{1/3} c2\ph s\th )$&	$36 i_{-5/3} s2\ph s2\th $\\
$K_{XXXZ}$ &	$12 s\th  (5 i_{-3/5} s\ph + 3 i_{-5/3} s\ph c2\th  - 6 i_{-5/3} s3\ph s^2\th)$&	$3 s\ph (11 i_{-7/11} - 3 i_{-5} c2\th ) c\th  + 9 i_{-5} s3\ph s2\th  s\th $&	$3 (-7 i_{-3/7} c\ph + 15 i_{1/3} c\ph c2\th  + 6 i_{-5} c3\ph s^2\th)$\\
$K_{XXYY}$ &	$144 i_{-5/3} s2\ph c\th  s^2\th$&	$-18 s2\ph (3 i_{-5/9} + i_{-5} c2\th ) s\th $&	$-36 i_{-5/3} c2\ph s2\th $\\
$K_{XXYZ}$ &	$-12 (-3 i_{-5/3} c\ph s3\th  + i_1 c\ph s\th  - 12 i_{-5/3} c3\ph s^3\th)$&	$6 c\th  (-5 i_{7/5} c\ph - 3 i_{-5} c\ph c2\th  - 6 i_{-5} c3\ph s^2\th)$&	$-6 (i_{3} s\ph - 9 i_{-5/9} s\ph c2\th  - 6 i_{-5} s3\ph s^2\th)$\\
$K_{XXZZ}$ &	$-144 i_{-5/3} s2\ph c\th  s^2\th$&	$-18 s2\ph (i_{5/3} - i_{-5} c2 \th ) s\th $&	$-12 (3 + 5 \iop c2\ph) s2\th $\\
$K_{XYYY}$ &	$-4 (3 i_{-5/3} c3\th  + 5 i_{-3/5} c\th  + 12 i_{-5/3} c2\ph c\th  s^2\th)$&	$-6 i_{-5} s^2\ph s3\th  - 7 i_{-3/7} s\th  + 15 i_{1/3} c2\ph s\th $&	$-12 i_{-5/3} s2\ph s2\th $\\
$K_{XYYZ}$ &	$12 s\th  (7 i_{-9/7} s\ph + 9 i_{-5/3} s\ph c2\th  + 6 i_{-5/3} s3\ph s^2\th)$&	$3 c\th  (s\ph (i_{-21} - 9 i_{-5} c2\th ) - 6 i_{-5} s3\ph s^2\th)$&	$-3 (5 i_{-9/5} c\ph + 3 i_{-5} c\ph c2\th  + 6 i_{-5} c3\ph s^2\th)$\\
$K_{XYZZ}$ &	$12 (3 i_{-5/3} c3\th  + i_{-3} c\th  + 12 i_{-5/3} c2\ph c\th  s^2\th)$&	$-3 (-6 i_{-5} s^2\ph s3\th  + i_{3} s\th  - 9 i_{-5/9} c2\ph s \th )$&	$-60 \iop s2\ph s2\th $\\
$K_{XZZZ}$ &	$-16 s\ph (i_{-3} + 3 i_{-5/3} c2\th ) s\th $&	$-4 s\ph (7 i_{-1}1 -  3 i_{-5} c2\th ) c\th $&	$-4 c\ph (i_{3} + 3 i_{5/3} c2\th )$\\
$K_{YXXZ}$ &	$12 s\th  (-7 i_{-9/7} c\ph - 9 i_{-5/3} c\ph c2\th  + 6 i_{-5/3} c3\ph s^2\th)$&	$-3 c\th  (c\ph (i_{-21} - 9 i_{-5} c2\th ) + 6 i_{-5} c3\ph s^2\th)$&	$3 (-5 i_{-9/5} s\ph - 3 i_{-5} s\ph c2\th  + 6 i_{-5} s3\ph s^2\th)$\\
$K_{YXYZ}$ &	$-24 s\th  (i_{-3} s\ph + 3 i_{-5/3} s\ph c2\th  - 6 i_{-5/3} s3\ph s^2\th)$&	$ 6s\ph c\th(5 i_{7/5} + 3 i_{-5} c2\th )  - 18 i_{-5} s3\ph s2\th  s\th $&	$-6 (i_{3} c\ph - 9 i_{-5/9} c\ph c2\th  + 6 i_{-5} c3\ph s^2\th)$\\
$K_{YXZZ}$ &	$-12 (3 i_{-5/3} c3\th  + i_{-3} c\th  - 12 i_{-5/3} c2\ph c\th  s^2\th)$&	$-3 (6 i_{-5} c^2\ph s3\th  - i_{3} s\th  - 9 i_{-5/9} c2\ph s\th )$&	$-60 \iop s2\ph s2\th $\\
$K_{YYYZ}$ &	$-12 s\th  (5 i_{-3/5} c\ph + 3 i_{-5/3} c\ph c2\th  + 6 i_{-5/3} c3\ph s^2\th)$&	$-3 c\th  (11 i_{-7/11} c\ph - 3 i_{-5} c\ph c2\th  - 6 i_{-5} c3\ph s^2\th)$&	$-3 (7 i_{-3/7} s\ph - 15 i_{1/3} s\ph c2\th + 6 i_{-5} s3\ph s^2\th)$\\
$K_{YYZZ}$ &	$144 i_{-5/3} s2\ph c\th  s^2\th$&	$6 s2\ph (3 i_{5/3} - 3 i_{-5} c2\th ) s\th $&	$12 (-3 + 5 \iop c2\ph) s2\th $\\
$K_{YZZZ}$ &	$16 c\ph (i_{-3} + 3 i_{-5/3} c2\th ) s\th $&	$4 c\ph (7 i_{-1}1 - 3 i_{-5} c2\th ) c\th $&	$-4 s\ph (i_{3} + 3 i_{5/3} c2\th )$\\
  \hline
\end{tabular}
\caption{
  The components of the quantity $\vec \Om^{(5)} $
  in spherical coordinates 
  are constructed by multiplying the expressions in each row under a given component with the corresponding coefficient
  for Lorentz violation appearing in the first column,
  summing over all rows,
  then multiplying the sum by the overall factor, $\frac{G M_\oplus}{32 R_\oplus^3}$.
  For example,
  $\Om^{(5)}_r = \frac{G M_\oplus}{32 R_\oplus^3}(-12 (3 i_{-5/3} c3\th  + 5 i_{-3/5} c\th  - 12 i_{-5/3} c2\ph c\th  s^2\th)K_{XXXY}+12 s\th  (5 i_{-3/5} s\ph + 3 i_{-5/3} s\ph c2\th  - 6 i_{-5/3} s3\ph s^2\th)K_{XXXZ} + \ldots)$.
  For brevity, trig functions are abbreviated with their first letter ($s=\sin$, $c=\cos$).
}\label{om5table}
\end{table*}

\begin{table*}
\label{tab:sme_coeff}
\begin{tabular}{|l|c|c|c|}
  \hline
  Coeff.	&$A_1$	 	&$A_2$	 	&	 $A_3$			\\
  \hline
 $K_{XXXY}$ &	$0$&	$36 i_{-5/3} s\al s\be s2\th $&	$0$\\
$K_{XXXZ}$ &	$3 s\al c\be c\th(11 i_{-7/11} - 3 i_{-5} c2\th )   + 12 c\al s\th(5 i_{-3/5} + 3 i_{-5/3} c2\th )  $&$0$&	$18 s^2\th (i_{-5} s\al c\be c\th  - 4 i_{-5/3} c\al s\th )$\\
$K_{XXYY}$ &	$0$&	$18 s\th  (8 i_{-5/3} c\al c\th  s\th -s\al c\be (3 i_{-5/9} + i_{-5} c2\th ))$&	$0$\\
$K_{XXYZ}$ &	$-6 s\al s\be (i_{3} - 9 i_{-5/9} c2\th )$&	$0$&	$36 i_{-5} s\al s\be s^2\th $\\
$K_{XXZZ}$ &	$0$&	$-18 s\th  (8 i_{-5/3} c\al c\th  s\th + s\al c\be (i_{5/3} - i_{-5} c2\th ) )$&	$0$\\
$K_{XYYY}$ &	$0$&	$-12 i_{-5/3} s\al s\be s2\th $&	$0$\\
$K_{XYYZ}$ &	$3 s\al c\be c\th (i_{-21} - 9 i_{-5} c2\th )   + 12 c\al s\th(7 i_{-9/7} + 9 i_{-5/3} c2\th ) $&	$0$&	$18 s^2\th (4 i_{-5/3} c\al s\th -i_{-5} s\al c\be c\th )$\\
$K_{XYZZ}$ &	$0$&	$-60 \iop s\al s\be s2\th$&	$0$\\
$K_{XZZZ}$ &	$-4 s\al c\be c\th(7 i_{-1} - 3 i_{-5} c2\th )   - 16 c\al s\th(i_{-3} + 3 i_{-5/3} c2\th ) $&	$0$&	$0$\\
$K_{YXXZ}$ &	$-3 s\al s\be (5 i_{-9/5} + 3 i_{-5} c2\th )$&	$0$&	$18 i_{-5} s\al s\be s^2\th $\\
$K_{YXYZ}$ &	$6 s\al c\be c\th (5 i_{7/5} + 3 i_{-5} c2\th ) + 12 c\al (-3 i_{-5/3} s3\th  + i_1 s\th )$&	$0$&	$36 s^2\th (4 i_{-5/3} c\al s\th -i_{-5} s\al c\be c\th )$\\
$K_{YXZZ}$ &	$0$&	$-60 \iop s\al s\be s2\th$&	$0$\\
$K_{YYYZ}$ &	$3 s\al s\be (-7 i_{-3/7} + 15 i_{1/3} c2\th )$&	$0$&	$-18 i_{-5} s\al s\be s^2\th $\\
$K_{YYZZ}$ &	$0$&	$6 s\th  (s\al c\be (3 i_{5/3} - 3 i_{-5} c2\th ) + 24 i_{-5/3} c\al c\th  s\th )$&	$0$\\
$K_{YZZZ}$ &	$-4 s\al s\be (i_{3} + 3 i_{5/3} c2\th )$&	$0$&	$0$\\
\hline
\end{tabular}
\caption{Amplitudes of $\sin n\ph$ harmonics of the beat frequency appearing in Eq.\ \rf{beat5}.
For brevity, trig functions are abbreviated with their first letter ($s=\sin$, $c=\cos$).
  The explicit form of each amplitude $A_n$ is constructed by
  multiplying the expressions in each row under the amplitude $A_n$ with the corresponding coefficient
  for Lorentz violation appearing in the first column
  and summing over all rows.
} \label{tab:sin5}
\end{table*}

\begin{table*}
\label{tab:sme_coeff}
\begin{tabular}{|l|c|c|c|}
  \hline
  Coeff.	& $B_1$ &$B_2$	 	&$B_3$	\\
  \hline
$K_{XXXY}$ &	$0$&	$  72 \
i_{-5/3} c\al s2\th  s\th -18 s\al c\be s\th (3 i_{-5/9} + i_{-5} c2\th ) $&	$0$\\
$K_{XXXZ}$ &	$3 s\al s\be (-7 i_{-3/7} + 15 i_{1/3} c2\th )$&	$0$&	$18 i_{-5} \
s\al s\be s^2\th $\\
$K_{XXYY}$ &	$0$&	$-36 i_{-5/3} s\al s\be s2\th $&	$0$\\
$K_{XXYZ}$ &	$-6 s\al c\be c\th (5 i_{7/5} + 3 i_{-5} c2\th ) + 24 c\al s\th (i_{-3} + 3 i_{-5/3} c2\th ) $&	$0$&	$36 s^2\th (4 i_{-5/3} c\al s\th -i_{-5} s\al c\be \
c\th   )$\\
$K_{XXZZ}$ &	$0$&	$-60 \iop s\al s\be s2\th $&	$0$\\
$K_{XYYY}$ &	$0$&	$6 s\th (s\al c\be (3 i_{-5/9} + i_{-5} c2\th ) - 4 i_{-5/3} c\al s2\th ) $&	$0$\\
$K_{XYYZ}$ &	$-3 s\al s\be (5 i_{-9/5} + 3 i_{-5} c2\th )$&	$0$&	$-18 i_{-5} \
s\al s\be s^2\th $\\
$K_{XYZZ}$ &	$0$&	$18 s\th (s\al c\be (i_{5/3} - i_{-5} c2\th ) + 4 i_{-5/3} c\al s2\th ) $&	$0$\\
$K_{XZZZ}$ &	$-4 s\al s\be (i_{3} + 3 i_{5/3} c2\th )$&	$0$&	$0$\\
$K_{YXXZ}$ &	$-3 s\al c\be c\th (i_{-21} - 9 i_{-5} c2\th ) - 12 c\al s\th (7 i_{-9/7} + 9 i_{-5/3} c2\th ) $&	$0$&	$18 s^2\th (4 i_{-5/3} c\al s\th -i_{-5} \
s\al c\be c\th   )$\\
$K_{YXYZ}$ &	$-6 s\al s\be (i_{3} - 9 i_{-5/9} c2\th )$&	$0$&	$-36 i_{-5} \
s\al s\be s^2\th $\\
$K_{YXZZ}$ &	$0$&	$18 s\th (s\al c\be (i_{5/3} - i_{-5} c2\th ) + 4 i_{-5/3} c\al s2\th ) $&	$0$\\
$K_{YYYZ}$ &	$3 s\al c\be c\th ( 3 i_{-5} c2\th -11 i_{-7/11}) - 6 c\al (3 \
i_{-5/3} s3\th  + 7 i_{-1/7} s\th )$&	$0$&	$18 s^2\th (i_{-5} \
s\al c\be c\th  - 4 i_{-5/3} c\al s\th )$\\
$K_{YYZZ}$ &	$0$&	$60 \iop s\al s\be s2\th$&	$0$\\
$K_{YZZZ}$ &	$4 s\al c\be c\th (7 i_{-1} - 3 i_{-5} c2\th ) + 16 c\al s\th (i_{-3} + 3 i_{-5/3} c2\th ) $&	$0$&	$0$\\
\hline
\end{tabular}
\caption{Amplitudes $B_n$ of $\cos n\ph$ harmonics of the beat frequency appearing in \rf{beat5}
  constructed as in Table \ref{tab:sin5}.
}\label{tab:cos5}
\end{table*}

\subsubsection{The minimal SME}

The leading minimal Lorentz-violating effects
on gyroscopes
are described by the 3 degrees of freedom
contained in $\sb^{TJ}$.
Due to its relative simplicity,
we apply the above results to the minimal SME first.
To obtain the dominant $\sb^\mn$ effects on the beat frequency
of laboratory gyroscopes,
it suffices to apply the above methods with the metric \rf{sme} expressed
in Earth-centered coordinates
in a spherical-Earth approximation,
where we find
that the dominant Lorentz-violating contributions to $\vec \Omega$ 
at $d=4$ can be written
\beq
\vec \Omega^{(4)} = \vec s \times \vec g,
\label{efom}
\eeq
where $\vec s = \stj$.
Transforming the metric to laboratory coordinates
would yield the usual Sagnac term
along with Lorentz-violating corrections to it.
Such suppressed corrections take us beyond Newtonian order
hence they are not considered here.
The Earth-centered components $\stj$
are equal to the Sun-centered frame components $\sb^{TJ}$
up to terms suppressed by the boost of the Earth on its orbit around the Sun,
a suppression factor of $10^{-4}$.

Note that $\vec \Om^{(4)}$,
which is proportional to the effective current density
in the Ampere's law analogy,
has no radial component.
Hence laser-gyroscope loops having $\hat n$ radial,
will involve no leading Lorentz-violation signal
from $\sb^{TJ}$.
The explicit form of the polar and azimuthal components
in Earth-centered spherical coordinates
offers some additional insights into the structure of the signal:
\bea
\Omega^{(4)}_\th &=& \fr{G M_\oplus}{R_\oplus^2} (\stx \sin \ph - \sty \cos \ph )\\
\Omega^{(4)}_\ph &=& \fr{G M_\oplus}{R_\oplus^2} \Big[\cos \th (\stx \cos \ph + \sty \sin \ph) - \stz \sin \th \Big].
\eea
The $\ph$ dependence of the signal
in a loop  oriented with $\hat n$
in the polar direction
is $90^\circ$ out of phase
with that of a loop oriented with $\hat n$
in the azimuthal direction,
and the latter orientation
is the only one with sensitivity to the $Z$ component of $\sb^{TJ}$.

Applying the procedure outlined above,
we find
\bea
\nonumber
f^{(4)}_b &=&
\fr{4 A G M_\oplus}{\la P R_\oplus^2} \sin \al
   \Big[ \cos \be ( \stx \sin \ph - \sty \cos \ph)\\
&&
 + \sin \be \left( \cos \th ( \stx \cos \ph + \sty \sin \ph) - \stz \sin \th\right) \Big] \pt{117}
\label{fblv}
\eea
for the explicit form of the leading $d=4$ contributions to the beat frequency in a laser-gyroscope system
of arbitrary orientation described by the angles $\al$ and $\be$
and arbitrary Earth-based location specified by the angles $\th$ and $\ph$.

In the context of matter-wave interferometers,
we also present the explicit form of the potentially useful combination
$\vec k \cdot (\vec g \times \vec \Om^{(4)})$.
To do so,
we write $\vec k$ in laboratory polar coordinates
again using $\al$ and $\be$ as the laboratory polar and azimuthal angles respectively,
and take $\vec g$ as radial:
\bea
\label{matter4}
\nonumber
\vec k \cdot (\vec g \times \vec \Om^{(4)})
&=&
\fr{k G^2 M^2_\oplus}{R_\oplus^4} \sin \al \Big[ \sin \be ( \sty \cos \ph - \stx \sin \ph)\\
  &&
  \nonumber
  + \cos \be \left( \cos \th ( \stx \cos \ph + \sty \sin \ph) - \stz \sin \th \right) \Big].\\
&& 
\eea
It should be noted that Lorentz-violating contributions to an effective $\vec g$
in the lab could, in principle, be of interest
and can arise at leading order in Lorentz violation
when multiplied by Lorentz-invariant contributions to $\vec \Om$.
In the context of our present focus on $\sb^{TJ}$ and $K_{JKLM}$
these contributions are suppressed relative to those
that arise from Eq.\ \rf{matter4}.

\subsubsection{Mass-dimension 5 signals}

The signals generated by Lorentz-violating operators of higher mass dimension
in gyroscope systems
can be generated by following the same procedures used at $d=4$
provided that the metric is available.
With the metric contributions at $d=5$ now available \cite{qblsvel}
as presented in Eq.\ \rf{sme},
we present the associated signals in interferometric gyroscope experiments.

The results are most neatly expressed by characterizing the 15 Lorentz-violating
degrees of freedom appearing in the $d=5$ signal in terms of the
15 canonical $K_{jklm}$ coefficients introduced in Ref.\ \cite{qblsvel}.
While the form of the signal at $d=5$ provides some experimental advantages,
the associated expressions are more lengthy.
The content of $\vec \Om^{(5)}$,
the $d=5$ contributions to $\vec \Om$,
is presented in Table \ref{om5table}.

\begin{table}
\label{tab:sme_coeff}
\begin{tabular}{|l|c|}
  \hline
  Coeff.	&$A_0$\\
  \hline
$K_{XXXY}$ &	$-12 s\al c\be s\th  (4 i_{-3} c^2\th + i_{3} s^2\th ) - 48 c\al c\th  (2 i_{-1} c^2\th - i_{-3} s^2\th )$\\
$K_{XXZZ}$ &	$-36 s\al s\be s2\th $\\
$K_{XYYY}$ &	$-4 s\al c\be s\th  (4 i_{-3} c^2\th + i_{3} s^2\th ) - 16 c\al c\th  (2 i_{-1} c^2\th - i_{-3} s^2\th )$\\
  $K_{XYZZ}$ &	$24 (2 i_{-2} c\al c^3\th + i_{-6} s\al c\be c^2\th s\th)  $\\
          &	$  -12( 4 i_{-3/2} c\al s2\th  s\th  + i_{-3} s\al c\be s^3\th )$\\
  $K_{YXZZ}$ &	$24 (2 i_{-2} c\al c^3\th + i_{-6} s\al c\be c^2\th s\th)  $\\
   &	$  -12( 4 i_{-3/2} c\al s2\th  s\th  + i_{-3} s\al c\be s^3\th )$\\
$K_{YYZZ}$ &	$-36 s\al s\be s2\th $\\
\hline
\end{tabular}
\caption{The time-independent contributions to the beat frequency appearing in \rf{beat5}
 constructed as in Table \ref{tab:sin5}.
}\label{tab:const5}
\end{table}

To present the key results in a way that
efficiently highlights trends, we
decompose the expressions by harmonics of the angle $\ph$,
and hence harmonics of the sidereal frequency.
The beat frequency in a laser-gyroscope system can be written
\beq
f^{(5)}_b =
\fr{A G M_\oplus}{8 \la P R_\oplus^3}\left(A_0 + \sum_{n=1}^3 (A_n \sin n\ph + B_n \cos n \ph)\right).
\label{beat5}
\eeq
The amplitudes $A_n$ and $B_n$ containing the coefficients for Lorentz violation
and the remaining angles describing the location and orientation of the ring are presented in Tables \ref{tab:sin5} - \ref{tab:const5}.
Entries in the tables take the form of experiment-specific numbers,
again highlighting the feature that measurements performed with different orientations
and in different Earth-based locations
will measure different linear combinations of coefficients for Lorentz violation.
Note that the signal at $d=5$ involves up to the 3rd harmonic of $\ph$.

Proceeding to provide the analogous information in the $d=5$ case,
the combination
$\vec k \cdot (\vec g \times \vec \Om^{(5)})$
takes the form
\beq
\vec k \cdot (\vec g \times \vec \Om^{(5)})
= \fr{k G^2 M^2_\oplus}{32 R_\oplus^5} \left(C_0 + \sum_{n=1}^3 (C_n \sin n\ph + D_n \cos n \ph)\right),
\label{matterfactor}
\eeq
where the amplitudes $C_n,D_n$
are provided in Tables \ref{tab:c} and \ref{tab:d}
using the same structure as in the case of $f^{(5)}_b$.

\begin{table*}
\label{tab:sme_coeff}
\begin{tabular}{|l|c|c|c|c|}
  \hline
  Coeff.     &$C_0$	&$C_1$	 	&$C_2$	 	&	 $C_3$			\\
  \hline
$K_{XXXY}$ &	$6 s\al s\be s\th (5 i_{-9/5} + 3 i_{-5} c2\th ) $&	$0$&	$36 i_{-5/3} s\al c\be s2\th $&	$0$\\
$K_{XXXZ}$ &	$0$&	$3 s\al s\be c\th (-11 i_{-7/11} + 3 i_{-5} c2\th ) $&	$0$&	$-9 \
i_{-5} s\al s\be s2\th  s\th $\\
$K_{XXYY}$ &	$0$&	$0$&	$18 s\al s\be s\th (3 i_{-5/9} + i_{-5} c2\th ) $&	$0$\\
$K_{XXYZ}$ &	$0$&	$-6 s\al c\be (i_{3} - 9 i_{-5/9} c2\th )$&	$0$&	$36 i_{-5} s\al \
c\be s^2\th $\\
$K_{XXZZ}$ &	$-72 s\al c\be c\th  s\th $&	$0$&	$18 s\al s\be s\th (i_{5/3} - i_{-5} c2\th ) $&	$0$\\
$K_{XYYY}$ &	$2 s\al s\be s\th (5 i_{-9/5} + 3 i_{-5} c2\th ) $&	$0$&	$-12 i_{-5/3} s\al c\be s2\th $&	$0$\\
$K_{XYYZ}$ &	$0$&	$3 s\al s\be c\th (-i_{-21} + 9 i_{-5} c2\th ) $&	$0$&	$9 i_{-5} \
s\al s\be s2\th  s\th $\\
$K_{XYZZ}$ &	$-6 s\al s\be s\th (i_{9} + 3 i_{-5} c2\th ) $&	$0$&	$-60 \iop s\al c\be s2\th $&	$0$\\
$K_{XZZZ}$ &	$0$&	$-4 s\al s\be c\th (-7 i_{-1} + 3 i_{-5} c2\th ) $&	$0$&	$0$\\
$K_{YXXZ}$ &	$0$&	$-3 s\al c\be (5 i_{-9/5} + 3 i_{-5} c2\th )$&	$0$&	$18 i_{-5} s\al c\be s^2\th $\\
$K_{YXYZ}$ &	$0$&	$-6 s\al s\be c\th (5 i_{7/5} + 3 i_{-5} c2\th ) $&	$0$&	$36 \
i_{-5} s\al s\be c\th  s^2\th $\\
$K_{YXZZ}$ &	$6 s\al s\be s\th (i_{9} + 3 i_{-5} c2\th ) $&	$0$&	$-60 \iop s\al c\be s2\th $&	$0$\\
$K_{YYYZ}$ &	$0$&	$3 s\al c\be (-7 i_{-3/7} + 15 i_{1/3} c2\th )$&	$0$&	$-18 \
i_{-5} s\al c\be s^2\th $\\
$K_{YYZZ}$ &	$-72 s\al c\be c\th  s\th $&	$0$&	$18 s\al s\be s\th (-i_{5/3} + i_{-5} c2\th ) $&	$0$\\
$K_{YZZZ}$ &	$0$&	$-4 s\al c\be (i_{3} + 3 i_{5/3} c2\th )$&	$0$&	$0$\\
  \hline
\end{tabular}
\caption{Amplitudes $C_n$ of $\sin n\ph$ harmonics of the matter-wave factor appearing in Eq.\ \rf{matterfactor},
  constructed as in Table \ref{tab:sin5}.
} \label{tab:c}
\end{table*}

\begin{table*}
\label{tab:sme_coeff}
\begin{tabular}{|l|c|c|c|}
  \hline
  Coeff.	&$D_1$	 	&$D_2$	 	&	 $D_3$			\\
  \hline
$K_{XXXY}$ &	$0$&	$18 s\al s\be s\th (3 i_{-5/9} + i_{-5} c2\th ) $&	$0$\\
$K_{XXXZ}$ &	$3 s\al c\be (-7 i_{-3/7} + 15 i_{1/3} c2\th )$&	$0$&	$18 i_{-5} \
s\al c\be s^2\th $\\
$K_{XXYY}$ &	$0$&	$-36 i_{-5/3} s\al c\be s2\th $&	$0$\\
$K_{XXYZ}$ &	$6 s\al s\be c\th (5 i_{7/5} + 3 i_{-5} c2\th ) $&	$0$&	$36 \
i_{-5} s\al s\be c\th  s^2\th $\\
$K_{XXZZ}$ &	$0$&	$-60 \iop s\al c\be s2\th$&	$0$\\
$K_{XYYY}$ &	$0$&	$-6 s\al s\be s\th (3 i_{-5/9} + i_{-5} c2\th ) $&	$0$\\
$K_{XYYZ}$ &	$-3 s\al c\be (5 i_{-9/5} + 3 i_{-5} c2\th )$&	$0$&	$-18 i_{-5} \
s\al c\be s^2\th $\\
$K_{XYZZ}$ &	$0$&	$18 s\al s\be s\th (-i_{5/3} + i_{-5} c2\th ) $&	$0$\\
$K_{XZZZ}$ &	$-4 s\al c\be (i_{3} + 3 i_{5/3} c2\th)$&	$0$&	$0$\\
$K_{YXXZ}$ &	$-3 s\al s\be c\th (-i_{-21} + 9 i_{-5} c2\th ) $&	$0$&	$9 \
i_{-5} s\al s\be s2\th  s\th $\\
$K_{YXYZ}$ &	$-6 s\al c\be (i_{3} - 9 i_{-5/9} c2\th )$&	$0$&	$-36 i_{-5} \
s\al c\be s^2\th $\\
$K_{YXZZ}$ &	$0$&	$18 s\al s\be s\th (-i_{5/3} + i_{-5} c2\th ) $&	$0$\\
$K_{YYYZ}$ &	$-3 s\al s\be c\th (-11 i_{-7/11} + 3 i_{-5} c2\th ) $&	$0$&	$-9 \
i_{-5} s\al s\be s2\th  s\th $\\
$K_{YYZZ}$ &	$0$&	$60 \iop s\al c\be s2\th$&	$0$\\
$K_{YZZZ}$ &	$4 s\al s\be c\th (-7 i_{-1} + 3 i_{-5} c2\th ) $&	$0$&	$0$\\
  \hline
\end{tabular}
\caption{Amplitudes $D_n$ of $\cos n\ph$ harmonics of the matter-wave factor appearing in Eq.\ \rf{matterfactor},
  constructed as in Table \ref{tab:sin5}.
} \label{tab:d}
\end{table*}

\section{Experiments and Measurements}
\label{expt}

In this section
we offer some additional comments
on translating experimental results
into measurements of the canonical forms of the coefficients for Lorentz violation.
We also discuss the sensitivities that existing devices
and those in development are likely to achieve.

In the present context of Earth-based experiments,
$\ph = \om_\oplus t + \ph_0$.
Here $\om_\oplus$ is the Earth's sidereal angular frequency,
$t$ is the local sidereal time,
and $\ph_0$ is a phase induced by the fact that $\ph$ as defined in this work
is not, in general, zero at the zero of local sidereal time.
This will be true only for experiments performed
at the longitude for which the Sun was directly over head
at the equator at the moment of the vernal equinox in the year 2000.
Based on data from the United States Naval Observatory \cite{usnavy},
this longitude $l_0 \approx 66.25^\circ$.
For experiments at other longitudes $l$ in degrees,
this makes \cite{akyd}
\beq
\ph_0 = \fr{2 \pi}{360} (l_0 - l).
\eeq

While Eqs.\ \rf{fblv} and \rf{beat5} can be fit to data from any 
laser-gyroscope system,
we point out a few features that may help identify a Lorentz-violation signal
among other signal and noise sources.
With the exception of special orientations,
the Lorentz-violation signal contains both constant effects
and effects periodic at the sidereal frequency
and its harmonics.
Such periodicity is not expected for the other signals
with perhaps the exception of tidal disturbances.
Moreover,
the conventional effects such as the dominant kinematic effect
of the rotating frame of the earth and gravitomagnetic effects do not generate signals for 
systems in which $\hat n$ is along $\hat \ph$
making the existence and form of a signal in a system
so oriented potentially of significant use.
In addition to these special features,
like the signals from conventional General-Relativistic effects,
the Lorentz-violation signal can be distinguished from the dominant kinematic effect
via comparison with sensitive earth-rotation measurements
from sources such as Very Long Baseline Interferometry.

The aim of devices under development is to measure general relativistic effects
that appear in these experiments analogous to a rotation of order $10^{-9} \om_\oplus$,
and sensitivity near this level has already been achieved \cite{beverini2016}.
Hence devices under development aim to exceed this level of sensitivity \cite{beverini2016, diVirgilio2017}.
Given the form of the effective rotation rate in Eq.\ \rf{efom},
sensitivities to the $\sb^\mn$ coefficients can be crudely estimated 
using SI units as follows:
\beq
\sb^{TJ} \approx \fr{\ep \om_\oplus c R^2_\oplus}{G M_\oplus}.
\label{sensitivity4}
\eeq 
Here $\ep$ is the fractional sensitivity to $\om_\oplus$
and $c$ is the speed of light.
With $\ep \approx 10^{-9}$,
Eq.\ \rf{sensitivity4} yields an estimated sensitivity to Lorentz violation of order $10^{-6}$.
This suggests advanced interferometric gyroscope experiments
are likely to be competitive with other laboratory \cite{flowers2016,chung2009,*muller2007}
and perhaps Solar-System experiments \cite{bourgoin2017} measuring $\sb^{TJ}$.
Though they are unlikely to compete with constraints set by astrophysical observations \cite{gwgrb2017,akjt15},
laboratory tests are often thought of differently due to the enhanced control
and understanding available to experiments over observations.

Though the specifics vary among the $d=5$ coefficients,
the dominant feature that appears when comparing with the estimated sensitivity to the minimal coefficients
is the additional factor of distance to the source in the beat frequency \rf{beat5}.
This can also be seen from the form of $\vec \Om^{(5)}$ presented in Table \ref{om5table},
where the coefficients for Lorentz violation enter $\vec \Om^{(5)}$
multiplied by $\frac{GM_\oplus}{R_\oplus^3}$ along with a numerical factor
of up to order 1.
Hence 
the $d=5$ sensitivities can be crudely estimated as
\beq
K_{JKLM} \approx \fr{\ep \om_\oplus c R^3_\oplus }{G M_\oplus},
\eeq
which suggests $d=5$ sensitivities better than 10 m are possible.
Sensitivities at this level would be competitive
with the best existing measurements of these coefficients,
which currently come from binary-pulsar analysis \cite{qblsvel}.

The capabilities of laser-gyroscope systems relative to other kinds of tests
can be understood from a combination of features.
First,
Lorentz violation entering the metric in $g_{0j}$
is typically suppressed in post-Newtonian tests by the relative speed of the source and test bodies.
In light-based gyroscopes, this issue is obviated by the use of light to probe the metric.
In comparison with the pulsar tests,
Earth-based gyroscope experiments amount to a shorter-range test,
which offers a relative advantage at $d=5$,
an advantage that could be expected to grow with mass dimension.
In fact,
interferometric gyroscopes may offer a novel system in which to search for
short-range velocity-dependent forces,
a qualitatively new type of signal in the search for new physics.

\section*{Acknowledgments}
The authors gratefully acknowledge the following for financial support:
S.M.\ from the Carleton College Summer Science Fellows Program
and
N.S.\ from the St.\ Olaf College CURI fund.
We also acknowledge useful conversations with 
D.\ Atkinson,
Q.\ Bailey,
A.\ Di Virgilio,
and M.\ Seifert.

\bibliography{cite.bib}

\begin{thebibliography}{52}%
\makeatletter
\providecommand \@ifxundefined [1]{%
 \@ifx{#1\undefined}
}%
\providecommand \@ifnum [1]{%
 \ifnum #1\expandafter \@firstoftwo
 \else \expandafter \@secondoftwo
 \fi
}%
\providecommand \@ifx [1]{%
 \ifx #1\expandafter \@firstoftwo
 \else \expandafter \@secondoftwo
 \fi
}%
\providecommand \natexlab [1]{#1}%
\providecommand \enquote  [1]{``#1''}%
\providecommand \bibnamefont  [1]{#1}%
\providecommand \bibfnamefont [1]{#1}%
\providecommand \citenamefont [1]{#1}%
\providecommand \href@noop [0]{\@secondoftwo}%
\providecommand \href [0]{\begingroup \@sanitize@url \@href}%
\providecommand \@href[1]{\@@startlink{#1}\@@href}%
\providecommand \@@href[1]{\endgroup#1\@@endlink}%
\providecommand \@sanitize@url [0]{\catcode `\\12\catcode `\$12\catcode
  `\&12\catcode `\#12\catcode `\^12\catcode `\_12\catcode `\%12\relax}%
\providecommand \@@startlink[1]{}%
\providecommand \@@endlink[0]{}%
\providecommand \url  [0]{\begingroup\@sanitize@url \@url }%
\providecommand \@url [1]{\endgroup\@href {#1}{\urlprefix }}%
\providecommand \urlprefix  [0]{URL }%
\providecommand \Eprint [0]{\href }%
\providecommand \doibase [0]{http://dx.doi.org/}%
\providecommand \selectlanguage [0]{\@gobble}%
\providecommand \bibinfo  [0]{\@secondoftwo}%
\providecommand \bibfield  [0]{\@secondoftwo}%
\providecommand \translation [1]{[#1]}%
\providecommand \BibitemOpen [0]{}%
\providecommand \bibitemStop [0]{}%
\providecommand \bibitemNoStop [0]{.\EOS\space}%
\providecommand \EOS [0]{\spacefactor3000\relax}%
\providecommand \BibitemShut  [1]{\csname bibitem#1\endcsname}%
\let\auto@bib@innerbib\@empty
%</preamble>
\bibitem [{\citenamefont {Sagnac}(1913)}]{sagnac1913}%
  \BibitemOpen
  \bibfield  {author} {\bibinfo {author} {\bibfnamefont {G.}~\bibnamefont
  {Sagnac}},\ }\href@noop {} {\bibfield  {journal} {\bibinfo  {journal} {C.R.
  Acad. Sci.}\ }\textbf {\bibinfo {volume} {157}},\ \bibinfo {pages} {708}
  (\bibinfo {year} {1913})}\BibitemShut {NoStop}%
\bibitem [{\citenamefont {Lawrence}(1998)}]{ring1}%
  \BibitemOpen
  \bibfield  {author} {\bibinfo {author} {\bibfnamefont {A.}~\bibnamefont
  {Lawrence}},\ }\href@noop {} {\emph {\bibinfo {title} {Modern Inertial
  Technology,}}}\ (\bibinfo  {publisher} {Springer, New York},\ \bibinfo {year}
  {1998})\BibitemShut {NoStop}%
\bibitem [{\citenamefont {Beverini}\ \emph {et~al.}(2016)\citenamefont
  {Beverini}, \citenamefont {Virgilio}, \citenamefont {Belfi}, \citenamefont
  {Ortolan}, \citenamefont {Schreiber}, \citenamefont {Gebauer},\ and\
  \citenamefont {Klügel}}]{beverini2016}%
  \BibitemOpen
  \bibfield  {author} {\bibinfo {author} {\bibfnamefont {N.}~\bibnamefont
  {Beverini}}, \bibinfo {author} {\bibfnamefont {A.~D.}\ \bibnamefont
  {Virgilio}}, \bibinfo {author} {\bibfnamefont {J.}~\bibnamefont {Belfi}},
  \bibinfo {author} {\bibfnamefont {A.}~\bibnamefont {Ortolan}}, \bibinfo
  {author} {\bibfnamefont {K.~U.}\ \bibnamefont {Schreiber}}, \bibinfo {author}
  {\bibfnamefont {A.}~\bibnamefont {Gebauer}}, \ and\ \bibinfo {author}
  {\bibfnamefont {T.}~\bibnamefont {Klügel}},\ }\href {\doibase
  10.1088/1742-6596/723/1/012061} {\bibfield  {journal} {\bibinfo  {journal}
  {J. Phys. Conf. Ser.}\ }\textbf {\bibinfo {volume} {723}},\ \bibinfo {pages}
  {012061} (\bibinfo {year} {2016})}\BibitemShut {NoStop}%
\bibitem [{\citenamefont {Bosi}\ \emph {et~al.}(2011)\citenamefont {Bosi},
  \citenamefont {Cella}, \citenamefont {Di~Virgilio}, \citenamefont {Ortolan},
  \citenamefont {Porzio}, \citenamefont {Solimeno}, \citenamefont {Cerdonio},
  \citenamefont {Zendri}, \citenamefont {Allegrini}, \citenamefont {Belfi}
  \emph {et~al.}}]{bosi2011}%
  \BibitemOpen
  \bibfield  {author} {\bibinfo {author} {\bibfnamefont {F.}~\bibnamefont
  {Bosi}}, \bibinfo {author} {\bibfnamefont {G.}~\bibnamefont {Cella}},
  \bibinfo {author} {\bibfnamefont {A.}~\bibnamefont {Di~Virgilio}}, \bibinfo
  {author} {\bibfnamefont {A.}~\bibnamefont {Ortolan}}, \bibinfo {author}
  {\bibfnamefont {A.}~\bibnamefont {Porzio}}, \bibinfo {author} {\bibfnamefont
  {S.}~\bibnamefont {Solimeno}}, \bibinfo {author} {\bibfnamefont
  {M.}~\bibnamefont {Cerdonio}}, \bibinfo {author} {\bibfnamefont {J.~P.}\
  \bibnamefont {Zendri}}, \bibinfo {author} {\bibfnamefont {M.}~\bibnamefont
  {Allegrini}}, \bibinfo {author} {\bibfnamefont {J.}~\bibnamefont {Belfi}},
  \emph {et~al.},\ }\href {\doibase 10.1103/PhysRevD.84.122002} {\bibfield
  {journal} {\bibinfo  {journal} {Phys. Rev. D}\ }\textbf {\bibinfo {volume}
  {84}},\ \bibinfo {pages} {122002} (\bibinfo {year} {2011})}\BibitemShut
  {NoStop}%
\bibitem [{\citenamefont {Storey}\ and\ \citenamefont
  {Cohen-Tannoudji}(1994)}]{ct}%
  \BibitemOpen
  \bibfield  {author} {\bibinfo {author} {\bibfnamefont {P.}~\bibnamefont
  {Storey}}\ and\ \bibinfo {author} {\bibfnamefont {C.}~\bibnamefont
  {Cohen-Tannoudji}},\ }\href {\doibase 10.1051/jp2:1994103} {\bibfield
  {journal} {\bibinfo  {journal} {J.\ Phys.\ II}\ }\textbf {\bibinfo {volume}
  {4}},\ \bibinfo {pages} {1999} (\bibinfo {year} {1994})}\BibitemShut
  {NoStop}%
\bibitem [{\citenamefont {Dubetsky}\ and\ \citenamefont
  {Kasevich}(2006)}]{dubetsky06}%
  \BibitemOpen
  \bibfield  {author} {\bibinfo {author} {\bibfnamefont {B.}~\bibnamefont
  {Dubetsky}}\ and\ \bibinfo {author} {\bibfnamefont {M.~A.}\ \bibnamefont
  {Kasevich}},\ }\href {\doibase 10.1103/PhysRevA.74.023615} {\bibfield
  {journal} {\bibinfo  {journal} {Phys. Rev. A}\ }\textbf {\bibinfo {volume}
  {74}},\ \bibinfo {pages} {023615} (\bibinfo {year} {2006})}\BibitemShut
  {NoStop}%
\bibitem [{\citenamefont {Colladay}\ and\ \citenamefont
  {Kosteleck\'y}(1998)}]{ck}%
  \BibitemOpen
  \bibfield  {author} {\bibinfo {author} {\bibfnamefont {D.}~\bibnamefont
  {Colladay}}\ and\ \bibinfo {author} {\bibfnamefont {V.~A.}\ \bibnamefont
  {Kosteleck\'y}},\ }\href {\doibase 10.1103/PhysRevD.58.116002} {\bibfield
  {journal} {\bibinfo  {journal} {\prd}\ }\textbf {\bibinfo {volume} {58}},\
  \bibinfo {pages} {116002} (\bibinfo {year} {1998})},\ \Eprint
  {http://arxiv.org/abs/hep-ph/9809521} {arXiv:hep-ph/9809521 [hep-ph]}
  \BibitemShut {NoStop}%
%%CITATION = HEP-PH/9809521;%%
\bibitem [{\citenamefont {Kosteleck\'y}(2004)}]{akgrav}%
  \BibitemOpen
  \bibfield  {author} {\bibinfo {author} {\bibfnamefont {V.~A.}\ \bibnamefont
  {Kosteleck\'y}},\ }\href {\doibase 10.1103/PhysRevD.69.105009} {\bibfield
  {journal} {\bibinfo  {journal} {\prd}\ }\textbf {\bibinfo {volume} {69}},\
  \bibinfo {pages} {105009} (\bibinfo {year} {2004})},\ \Eprint
  {http://arxiv.org/abs/hep-th/0312310} {arXiv:hep-th/0312310 [hep-th]}
  \BibitemShut {NoStop}%
%%CITATION = HEP-TH/0312310;%%
\bibitem [{\citenamefont {Ruggiero}(2015)}]{ruggiero15}%
  \BibitemOpen
  \bibfield  {author} {\bibinfo {author} {\bibfnamefont {M.~L.}\ \bibnamefont
  {Ruggiero}},\ }\href {\doibase 10.3390/galaxies3020084} {\bibfield  {journal}
  {\bibinfo  {journal} {Galaxies}\ }\textbf {\bibinfo {volume} {3}},\ \bibinfo
  {pages} {84} (\bibinfo {year} {2015})},\ \Eprint
  {http://arxiv.org/abs/1505.01268} {arXiv:1505.01268 [gr-qc]} \BibitemShut
  {NoStop}%
%%CITATION = ARXIV:1505.01268;%%
\bibitem [{\citenamefont {Scaramuzza}\ and\ \citenamefont
  {Tasson}(2017)}]{scaramuzza16}%
  \BibitemOpen
  \bibfield  {author} {\bibinfo {author} {\bibfnamefont {N.}~\bibnamefont
  {Scaramuzza}}\ and\ \bibinfo {author} {\bibfnamefont {J.~D.}\ \bibnamefont
  {Tasson}},\ }in\ \href {\doibase 10.1142/9789813148505_0072} {\emph {\bibinfo
  {booktitle} {{CPT and Lorentz Symmetry VII}}}},\ \bibinfo {editor} {edited
  by\ \bibinfo {editor} {\bibfnamefont {V.}~\bibnamefont {Kosteleck\'y}}}\
  (\bibinfo {year} {2017})\ pp.\ \bibinfo {pages} {271--273},\ \Eprint
  {http://arxiv.org/abs/1607.08111} {arXiv:1607.08111 [gr-qc]} \BibitemShut
  {NoStop}%
%%CITATION = ARXIV:1607.08111;%%
\bibitem [{\citenamefont {Kosteleck\'y}\ and\ \citenamefont
  {Samuel}(1989)}]{ksp}%
  \BibitemOpen
  \bibfield  {author} {\bibinfo {author} {\bibfnamefont {V.~A.}\ \bibnamefont
  {Kosteleck\'y}}\ and\ \bibinfo {author} {\bibfnamefont {S.}~\bibnamefont
  {Samuel}},\ }\href@noop {} {\bibfield  {journal} {\bibinfo  {journal} {Phys.
  Rev. D}\ }\textbf {\bibinfo {volume} {39}},\ \bibinfo {pages} {683} (\bibinfo
  {year} {1989})}\BibitemShut {NoStop}%
\bibitem [{\citenamefont {{Kosteleck{\'y}}}\ and\ \citenamefont
  {{Potting}}(1991)}]{akrp1991}%
  \BibitemOpen
  \bibfield  {author} {\bibinfo {author} {\bibfnamefont {V.~A.}\ \bibnamefont
  {{Kosteleck{\'y}}}}\ and\ \bibinfo {author} {\bibfnamefont {R.}~\bibnamefont
  {{Potting}}},\ }\href {\doibase 10.1016/0550-3213(91)90071-5} {\bibfield
  {journal} {\bibinfo  {journal} {Nucl. Phys. B}\ }\textbf {\bibinfo {volume}
  {359}},\ \bibinfo {pages} {545} (\bibinfo {year} {1991})}\BibitemShut
  {NoStop}%
\bibitem [{\citenamefont {Tasson}(2014)}]{tasson14}%
  \BibitemOpen
  \bibfield  {author} {\bibinfo {author} {\bibfnamefont {J.~D.}\ \bibnamefont
  {Tasson}},\ }\href {\doibase 10.1088/0034-4885/77/6/062901} {\bibfield
  {journal} {\bibinfo  {journal} {Rep. Prog. Phys.}\ }\textbf {\bibinfo
  {volume} {77}},\ \bibinfo {pages} {062901} (\bibinfo {year} {2014})},\
  \Eprint {http://arxiv.org/abs/1403.7785} {arXiv:1403.7785 [hep-ph]}
  \BibitemShut {NoStop}%
%%CITATION = ARXIV:1403.7785;%%
\bibitem [{\citenamefont {Bertschinger}\ \emph {et~al.}(2019)\citenamefont
  {Bertschinger}, \citenamefont {Flowers}, \citenamefont {Moseley},
  \citenamefont {Pfeifer}, \citenamefont {Tasson},\ and\ \citenamefont
  {Yang}}]{bertschinger18}%
  \BibitemOpen
  \bibfield  {author} {\bibinfo {author} {\bibfnamefont {T.~H.}\ \bibnamefont
  {Bertschinger}}, \bibinfo {author} {\bibfnamefont {N.~A.}\ \bibnamefont
  {Flowers}}, \bibinfo {author} {\bibfnamefont {S.}~\bibnamefont {Moseley}},
  \bibinfo {author} {\bibfnamefont {C.~R.}\ \bibnamefont {Pfeifer}}, \bibinfo
  {author} {\bibfnamefont {J.~D.}\ \bibnamefont {Tasson}}, \ and\ \bibinfo
  {author} {\bibfnamefont {S.}~\bibnamefont {Yang}},\ }\href {\doibase
  10.3390/sym11010022} {\bibfield  {journal} {\bibinfo  {journal} {Symmetry}\
  }\textbf {\bibinfo {volume} {11}},\ \bibinfo {pages} {22} (\bibinfo {year}
  {2019})}\BibitemShut {NoStop}%
%%CITATION = 00762,11,22;%%
\bibitem [{\citenamefont {Bailey}\ and\ \citenamefont {Lane}(2018)}]{qbcl2018}%
  \BibitemOpen
  \bibfield  {author} {\bibinfo {author} {\bibfnamefont {Q.~G.}\ \bibnamefont
  {Bailey}}\ and\ \bibinfo {author} {\bibfnamefont {C.~D.}\ \bibnamefont
  {Lane}},\ }\href {\doibase 10.3390/sym10100480} {\bibfield  {journal}
  {\bibinfo  {journal} {Symmetry}\ }\textbf {\bibinfo {volume} {10}},\ \bibinfo
  {pages} {480} (\bibinfo {year} {2018})},\ \Eprint
  {http://arxiv.org/abs/1810.05136} {arXiv:1810.05136 [hep-th]} \BibitemShut
  {NoStop}%
%%CITATION = ARXIV:1810.05136;%%
\bibitem [{\citenamefont {Kostelecký}\ and\ \citenamefont
  {Li}(2019)}]{akzl2019}%
  \BibitemOpen
  \bibfield  {author} {\bibinfo {author} {\bibfnamefont {V.~A.}\ \bibnamefont
  {Kostelecký}}\ and\ \bibinfo {author} {\bibfnamefont {Z.}~\bibnamefont
  {Li}},\ }\href {\doibase 10.1103/PhysRevD.99.056016} {\bibfield  {journal}
  {\bibinfo  {journal} {Phys. Rev. D}\ }\textbf {\bibinfo {volume} {99}},\
  \bibinfo {pages} {056016} (\bibinfo {year} {2019})},\ \Eprint
  {http://arxiv.org/abs/1812.11672} {arXiv:1812.11672 [hep-ph]} \BibitemShut
  {NoStop}%
%%CITATION = ARXIV:1812.11672;%%
\bibitem [{\citenamefont {Edwards}\ and\ \citenamefont
  {Kosteleck\'y}(2018)}]{akbe2018}%
  \BibitemOpen
  \bibfield  {author} {\bibinfo {author} {\bibfnamefont {B.~R.}\ \bibnamefont
  {Edwards}}\ and\ \bibinfo {author} {\bibfnamefont {V.~A.}\ \bibnamefont
  {Kosteleck\'y}},\ }\href {\doibase 10.1016/j.physletb.2018.10.011} {\bibfield
   {journal} {\bibinfo  {journal} {Phys. Lett. B}\ }\textbf {\bibinfo {volume}
  {786}},\ \bibinfo {pages} {319} (\bibinfo {year} {2018})},\ \Eprint
  {http://arxiv.org/abs/1809.05535} {arXiv:1809.05535 [hep-th]} \BibitemShut
  {NoStop}%
%%CITATION = ARXIV:1809.05535;%%
\bibitem [{\citenamefont {Kostelecký}\ and\ \citenamefont
  {Vargas}(2015)}]{akav2015}%
  \BibitemOpen
  \bibfield  {author} {\bibinfo {author} {\bibfnamefont {V.~A.}\ \bibnamefont
  {Kostelecký}}\ and\ \bibinfo {author} {\bibfnamefont {A.~J.}\ \bibnamefont
  {Vargas}},\ }\href {\doibase 10.1103/PhysRevD.92.056002} {\bibfield
  {journal} {\bibinfo  {journal} {Phys. Rev. D}\ }\textbf {\bibinfo {volume}
  {92}},\ \bibinfo {pages} {056002} (\bibinfo {year} {2015})},\ \Eprint
  {http://arxiv.org/abs/1506.01706} {arXiv:1506.01706 [hep-ph]} \BibitemShut
  {NoStop}%
%%CITATION = ARXIV:1506.01706;%%
\bibitem [{\citenamefont {Kosteleck\'y}\ and\ \citenamefont
  {Lehnert}(2001)}]{akrl2001}%
  \BibitemOpen
  \bibfield  {author} {\bibinfo {author} {\bibfnamefont {V.~A.}\ \bibnamefont
  {Kosteleck\'y}}\ and\ \bibinfo {author} {\bibfnamefont {R.}~\bibnamefont
  {Lehnert}},\ }\href {\doibase 10.1103/PhysRevD.63.065008} {\bibfield
  {journal} {\bibinfo  {journal} {Phys. Rev. D}\ }\textbf {\bibinfo {volume}
  {63}},\ \bibinfo {pages} {065008} (\bibinfo {year} {2001})},\ \Eprint
  {http://arxiv.org/abs/hep-th/0012060} {arXiv:hep-th/0012060 [hep-th]}
  \BibitemShut {NoStop}%
%%CITATION = HEP-TH/0012060;%%
\bibitem [{\citenamefont {Bonder}\ and\ \citenamefont
  {Corral}(2018)}]{bonder2018}%
  \BibitemOpen
  \bibfield  {author} {\bibinfo {author} {\bibfnamefont {Y.}~\bibnamefont
  {Bonder}}\ and\ \bibinfo {author} {\bibfnamefont {C.}~\bibnamefont
  {Corral}},\ }\href {\doibase 10.3390/sym10100433} {\bibfield  {journal}
  {\bibinfo  {journal} {Symmetry}\ }\textbf {\bibinfo {volume} {10}},\ \bibinfo
  {pages} {433} (\bibinfo {year} {2018})},\ \Eprint
  {http://arxiv.org/abs/1808.05522} {arXiv:1808.05522 [gr-qc]} \BibitemShut
  {NoStop}%
%%CITATION = ARXIV:1808.05522;%%
\bibitem [{\citenamefont {Seifert}(2018)}]{seifert2018}%
  \BibitemOpen
  \bibfield  {author} {\bibinfo {author} {\bibfnamefont {M.}~\bibnamefont
  {Seifert}},\ }\href {\doibase 10.3390/sym10100490} {\bibfield  {journal}
  {\bibinfo  {journal} {Symmetry}\ }\textbf {\bibinfo {volume} {10}},\ \bibinfo
  {pages} {490} (\bibinfo {year} {2018})}\BibitemShut {NoStop}%
%%CITATION = 00762,10,490;%%
\bibitem [{\citenamefont {Bluhm}\ and\ \citenamefont
  {Sehic}(2016)}]{bluhm2016}%
  \BibitemOpen
  \bibfield  {author} {\bibinfo {author} {\bibfnamefont {R.}~\bibnamefont
  {Bluhm}}\ and\ \bibinfo {author} {\bibfnamefont {A.}~\bibnamefont {Sehic}},\
  }\href {\doibase 10.1103/PhysRevD.94.104034} {\bibfield  {journal} {\bibinfo
  {journal} {Phys. Rev. D}\ }\textbf {\bibinfo {volume} {94}},\ \bibinfo
  {pages} {104034} (\bibinfo {year} {2016})},\ \Eprint
  {http://arxiv.org/abs/1610.02892} {arXiv:1610.02892 [hep-th]} \BibitemShut
  {NoStop}%
%%CITATION = ARXIV:1610.02892;%%
\bibitem [{\citenamefont {{Kosteleck{\'y}}}\ and\ \citenamefont
  {{Russell}}()}]{data}%
  \BibitemOpen
  \bibfield  {author} {\bibinfo {author} {\bibfnamefont {V.~A.}\ \bibnamefont
  {{Kosteleck{\'y}}}}\ and\ \bibinfo {author} {\bibfnamefont {N.}~\bibnamefont
  {{Russell}}},\ }\href@noop {} {\bibfield  {journal} {\bibinfo  {journal}
  {{\it Data tables for Lorentz and CPT violation}, 2018 Edition,}\ }}\Eprint
  {http://arxiv.org/abs/0801.0287v12} {arXiv:0801.0287v12 [hep-ph]}
  \BibitemShut {NoStop}%
\bibitem [{\citenamefont {{Bailey}}\ and\ \citenamefont
  {{Kosteleck{\'y}}}(2006)}]{lvpn}%
  \BibitemOpen
  \bibfield  {author} {\bibinfo {author} {\bibfnamefont {Q.~G.}\ \bibnamefont
  {{Bailey}}}\ and\ \bibinfo {author} {\bibfnamefont {V.~A.}\ \bibnamefont
  {{Kosteleck{\'y}}}},\ }\href {\doibase 10.1103/PhysRevD.74.045001} {\bibfield
   {journal} {\bibinfo  {journal} {\prd}\ }\textbf {\bibinfo {volume} {74}},\
  \bibinfo {eid} {045001} (\bibinfo {year} {2006})},\ \Eprint
  {http://arxiv.org/abs/gr-qc/0603030} {arXiv:gr-qc/0603030 [gr-qc]}
  \BibitemShut {NoStop}%
\bibitem [{\citenamefont {{Kosteleck{\'y}}}\ and\ \citenamefont
  {{Tasson}}(2011)}]{lvgap}%
  \BibitemOpen
  \bibfield  {author} {\bibinfo {author} {\bibfnamefont {V.~A.}\ \bibnamefont
  {{Kosteleck{\'y}}}}\ and\ \bibinfo {author} {\bibfnamefont {J.~D.}\
  \bibnamefont {{Tasson}}},\ }\href {\doibase 10.1103/PhysRevD.83.016013}
  {\bibfield  {journal} {\bibinfo  {journal} {\prd}\ }\textbf {\bibinfo
  {volume} {83}},\ \bibinfo {eid} {016013} (\bibinfo {year} {2011})},\ \Eprint
  {http://arxiv.org/abs/1006.4106} {arXiv:1006.4106 [gr-qc]} \BibitemShut
  {NoStop}%
\bibitem [{\citenamefont {Kosteleck{\'{y}}}\ and\ \citenamefont
  {Mewes}(2016)}]{kmgw}%
  \BibitemOpen
  \bibfield  {author} {\bibinfo {author} {\bibfnamefont {V.~A.}\ \bibnamefont
  {Kosteleck{\'{y}}}}\ and\ \bibinfo {author} {\bibfnamefont {M.}~\bibnamefont
  {Mewes}},\ }\href {\doibase 10.1016/j.physletb.2016.04.040} {\bibfield
  {journal} {\bibinfo  {journal} {Phys. Lett. B}\ }\textbf {\bibinfo {volume}
  {757}},\ \bibinfo {pages} {510} (\bibinfo {year} {2016})}\BibitemShut
  {NoStop}%
\bibitem [{\citenamefont {Kostelecký}\ and\ \citenamefont
  {Mewes}(2017)}]{kmSR2016}%
  \BibitemOpen
  \bibfield  {author} {\bibinfo {author} {\bibfnamefont {V.~A.}\ \bibnamefont
  {Kostelecký}}\ and\ \bibinfo {author} {\bibfnamefont {M.}~\bibnamefont
  {Mewes}},\ }\href {\doibase 10.1016/j.physletb.2016.12.062} {\bibfield
  {journal} {\bibinfo  {journal} {Phys. Lett. B}\ }\textbf {\bibinfo {volume}
  {766}},\ \bibinfo {pages} {137} (\bibinfo {year} {2017})},\ \Eprint
  {http://arxiv.org/abs/1611.10313} {arXiv:1611.10313 [gr-qc]} \BibitemShut
  {NoStop}%
%%CITATION = ARXIV:1611.10313;%%
\bibitem [{\citenamefont {Kostelecký}\ and\ \citenamefont
  {Mewes}(2018)}]{kmLG2017}%
  \BibitemOpen
  \bibfield  {author} {\bibinfo {author} {\bibfnamefont {V.~A.}\ \bibnamefont
  {Kostelecký}}\ and\ \bibinfo {author} {\bibfnamefont {M.}~\bibnamefont
  {Mewes}},\ }\href {\doibase 10.1016/j.physletb.2018.01.082} {\bibfield
  {journal} {\bibinfo  {journal} {Phys. Lett. B}\ }\textbf {\bibinfo {volume}
  {779}},\ \bibinfo {pages} {136} (\bibinfo {year} {2018})},\ \Eprint
  {http://arxiv.org/abs/1712.10268} {arXiv:1712.10268 [gr-qc]} \BibitemShut
  {NoStop}%
%%CITATION = ARXIV:1712.10268;%%
\bibitem [{\citenamefont {Bailey}\ \emph {et~al.}(2015)\citenamefont {Bailey},
  \citenamefont {Kostelecký},\ and\ \citenamefont {Xu}}]{qbakrx2014}%
  \BibitemOpen
  \bibfield  {author} {\bibinfo {author} {\bibfnamefont {Q.~G.}\ \bibnamefont
  {Bailey}}, \bibinfo {author} {\bibfnamefont {A.}~\bibnamefont {Kostelecký}},
  \ and\ \bibinfo {author} {\bibfnamefont {R.}~\bibnamefont {Xu}},\ }\href
  {\doibase 10.1103/PhysRevD.91.022006} {\bibfield  {journal} {\bibinfo
  {journal} {Phys. Rev. D}\ }\textbf {\bibinfo {volume} {91}},\ \bibinfo
  {pages} {022006} (\bibinfo {year} {2015})},\ \Eprint
  {http://arxiv.org/abs/1410.6162} {arXiv:1410.6162 [gr-qc]} \BibitemShut
  {NoStop}%
%%CITATION = ARXIV:1410.6162;%%
\bibitem [{\citenamefont {Mewes}(2019)}]{mewes2019}%
  \BibitemOpen
  \bibfield  {author} {\bibinfo {author} {\bibfnamefont {M.}~\bibnamefont
  {Mewes}},\ }\href {\doibase 10.1103/PhysRevD.99.104062} {\bibfield  {journal}
  {\bibinfo  {journal} {Phys. Rev. D}\ }\textbf {\bibinfo {volume} {99}},\
  \bibinfo {pages} {104062} (\bibinfo {year} {2019})},\ \Eprint
  {http://arxiv.org/abs/1905.00409} {arXiv:1905.00409 [gr-qc]} \BibitemShut
  {NoStop}%
%%CITATION = ARXIV:1905.00409;%%
\bibitem [{\citenamefont {Bailey}\ and\ \citenamefont {Havert}(2017)}]{qd5}%
  \BibitemOpen
  \bibfield  {author} {\bibinfo {author} {\bibfnamefont {Q.~G.}\ \bibnamefont
  {Bailey}}\ and\ \bibinfo {author} {\bibfnamefont {D.}~\bibnamefont
  {Havert}},\ }\href {\doibase 10.1103/PhysRevD.96.064035} {\bibfield
  {journal} {\bibinfo  {journal} {Phys. Rev. D}\ }\textbf {\bibinfo {volume}
  {96}},\ \bibinfo {pages} {064035} (\bibinfo {year} {2017})}\BibitemShut
  {NoStop}%
\bibitem [{\citenamefont {Shao}\ and\ \citenamefont {Bailey}(2018)}]{qblsvel}%
  \BibitemOpen
  \bibfield  {author} {\bibinfo {author} {\bibfnamefont {L.}~\bibnamefont
  {Shao}}\ and\ \bibinfo {author} {\bibfnamefont {Q.~G.}\ \bibnamefont
  {Bailey}},\ }\href {\doibase 10.1103/PhysRevD.98.084049} {\bibfield
  {journal} {\bibinfo  {journal} {Phys. Rev. D}\ }\textbf {\bibinfo {volume}
  {98}},\ \bibinfo {pages} {084049} (\bibinfo {year} {2018})}\BibitemShut
  {NoStop}%
\bibitem [{\citenamefont {Shao}\ and\ \citenamefont
  {Bailey}(2019)}]{shaol2019}%
  \BibitemOpen
  \bibfield  {author} {\bibinfo {author} {\bibfnamefont {L.}~\bibnamefont
  {Shao}}\ and\ \bibinfo {author} {\bibfnamefont {Q.~G.}\ \bibnamefont
  {Bailey}},\ }\href {\doibase 10.1103/PhysRevD.99.084017} {\bibfield
  {journal} {\bibinfo  {journal} {Phys. Rev. D}\ }\textbf {\bibinfo {volume}
  {99}},\ \bibinfo {pages} {084017} (\bibinfo {year} {2019})},\ \Eprint
  {http://arxiv.org/abs/1903.11760} {arXiv:1903.11760 [gr-qc]} \BibitemShut
  {NoStop}%
%%CITATION = ARXIV:1903.11760;%%
\bibitem [{\citenamefont {Shao}\ \emph {et~al.}(2019)\citenamefont {Shao},
  \citenamefont {Chen}, \citenamefont {Tan}, \citenamefont {Yang},
  \citenamefont {Luo}, \citenamefont {Tobar}, \citenamefont {Long},
  \citenamefont {Weisman},\ and\ \citenamefont {Kosteleck\'y}}]{cgshao2019}%
  \BibitemOpen
  \bibfield  {author} {\bibinfo {author} {\bibfnamefont {C.-G.}\ \bibnamefont
  {Shao}}, \bibinfo {author} {\bibfnamefont {Y.-F.}\ \bibnamefont {Chen}},
  \bibinfo {author} {\bibfnamefont {Y.-J.}\ \bibnamefont {Tan}}, \bibinfo
  {author} {\bibfnamefont {S.-Q.}\ \bibnamefont {Yang}}, \bibinfo {author}
  {\bibfnamefont {J.}~\bibnamefont {Luo}}, \bibinfo {author} {\bibfnamefont
  {M.~E.}\ \bibnamefont {Tobar}}, \bibinfo {author} {\bibfnamefont {J.~C.}\
  \bibnamefont {Long}}, \bibinfo {author} {\bibfnamefont {E.}~\bibnamefont
  {Weisman}}, \ and\ \bibinfo {author} {\bibfnamefont {V.~A.}\ \bibnamefont
  {Kosteleck\'y}},\ }\href {\doibase 10.1103/PhysRevLett.122.011102} {\bibfield
   {journal} {\bibinfo  {journal} {Phys. Rev. Lett.}\ }\textbf {\bibinfo
  {volume} {122}},\ \bibinfo {pages} {011102} (\bibinfo {year}
  {2019})}\BibitemShut {NoStop}%
\bibitem [{\citenamefont {Abbott}\ \emph {et~al.}(2017)\citenamefont {Abbott}
  \emph {et~al.}}]{gwgrb2017}%
  \BibitemOpen
  \bibfield  {author} {\bibinfo {author} {\bibfnamefont {B.~P.}\ \bibnamefont
  {Abbott}} \emph {et~al.} (\bibinfo {collaboration} {LIGO Scientific, Virgo,
  Fermi-GBM, and INTEGRAL Collaborations}),\ }\href {\doibase
  10.3847/2041-8213/aa920c} {\bibfield  {journal} {\bibinfo  {journal}
  {Astrophys. J.}\ }\textbf {\bibinfo {volume} {848}},\ \bibinfo {pages} {L13}
  (\bibinfo {year} {2017})},\ \Eprint {http://arxiv.org/abs/1710.05834}
  {arXiv:1710.05834 [astro-ph.HE]} \BibitemShut {NoStop}%
%%CITATION = ARXIV:1710.05834;%%
\bibitem [{\citenamefont {Shao}\ \emph {et~al.}(2018)\citenamefont {Shao},
  \citenamefont {Chen}, \citenamefont {Sun}, \citenamefont {Cao}, \citenamefont
  {Zhou}, \citenamefont {Hu}, \citenamefont {Yu},\ and\ \citenamefont
  {Müller}}]{shaoGmeter2017}%
  \BibitemOpen
  \bibfield  {author} {\bibinfo {author} {\bibfnamefont {C.-G.}\ \bibnamefont
  {Shao}}, \bibinfo {author} {\bibfnamefont {Y.-F.}\ \bibnamefont {Chen}},
  \bibinfo {author} {\bibfnamefont {R.}~\bibnamefont {Sun}}, \bibinfo {author}
  {\bibfnamefont {L.-S.}\ \bibnamefont {Cao}}, \bibinfo {author} {\bibfnamefont
  {M.-K.}\ \bibnamefont {Zhou}}, \bibinfo {author} {\bibfnamefont {Z.-K.}\
  \bibnamefont {Hu}}, \bibinfo {author} {\bibfnamefont {C.}~\bibnamefont {Yu}},
  \ and\ \bibinfo {author} {\bibfnamefont {H.}~\bibnamefont {Müller}},\ }\href
  {\doibase 10.1103/PhysRevD.97.024019} {\bibfield  {journal} {\bibinfo
  {journal} {Phys. Rev. D}\ }\textbf {\bibinfo {volume} {97}},\ \bibinfo
  {pages} {024019} (\bibinfo {year} {2018})},\ \Eprint
  {http://arxiv.org/abs/1707.02318} {arXiv:1707.02318 [gr-qc]} \BibitemShut
  {NoStop}%
%%CITATION = ARXIV:1707.02318;%%
\bibitem [{\citenamefont {Bourgoin}\ \emph {et~al.}(2017)\citenamefont
  {Bourgoin}, \citenamefont {Le~Poncin-Lafitte}, \citenamefont {Hees},
  \citenamefont {Bouquillon}, \citenamefont {Francou},\ and\ \citenamefont
  {Angonin}}]{bourgoin2017}%
  \BibitemOpen
  \bibfield  {author} {\bibinfo {author} {\bibfnamefont {A.}~\bibnamefont
  {Bourgoin}}, \bibinfo {author} {\bibfnamefont {C.}~\bibnamefont
  {Le~Poncin-Lafitte}}, \bibinfo {author} {\bibfnamefont {A.}~\bibnamefont
  {Hees}}, \bibinfo {author} {\bibfnamefont {S.}~\bibnamefont {Bouquillon}},
  \bibinfo {author} {\bibfnamefont {G.}~\bibnamefont {Francou}}, \ and\
  \bibinfo {author} {\bibfnamefont {M.-C.}\ \bibnamefont {Angonin}},\ }\href
  {\doibase 10.1103/PhysRevLett.119.201102} {\bibfield  {journal} {\bibinfo
  {journal} {Phys. Rev. Lett.}\ }\textbf {\bibinfo {volume} {119}},\ \bibinfo
  {pages} {201102} (\bibinfo {year} {2017})},\ \Eprint
  {http://arxiv.org/abs/1706.06294} {arXiv:1706.06294 [gr-qc]} \BibitemShut
  {NoStop}%
%%CITATION = ARXIV:1706.06294;%%
\bibitem [{\citenamefont {Flowers}\ \emph {et~al.}(2017)\citenamefont
  {Flowers}, \citenamefont {Goodge},\ and\ \citenamefont
  {Tasson}}]{flowers2016}%
  \BibitemOpen
  \bibfield  {author} {\bibinfo {author} {\bibfnamefont {N.~A.}\ \bibnamefont
  {Flowers}}, \bibinfo {author} {\bibfnamefont {C.}~\bibnamefont {Goodge}}, \
  and\ \bibinfo {author} {\bibfnamefont {J.~D.}\ \bibnamefont {Tasson}},\
  }\href {\doibase 10.1103/PhysRevLett.119.201101} {\bibfield  {journal}
  {\bibinfo  {journal} {Phys. Rev. Lett.}\ }\textbf {\bibinfo {volume} {119}},\
  \bibinfo {pages} {201101} (\bibinfo {year} {2017})},\ \Eprint
  {http://arxiv.org/abs/1612.08495} {arXiv:1612.08495 [gr-qc]} \BibitemShut
  {NoStop}%
%%CITATION = ARXIV:1612.08495;%%
\bibitem [{\citenamefont {Tasson}(2017)}]{tasson17}%
  \BibitemOpen
  \bibfield  {author} {\bibinfo {author} {\bibfnamefont {J.~D.}\ \bibnamefont
  {Tasson}},\ }\href {\doibase 10.7566/JPSCP.18.011002} {\bibfield  {journal}
  {\bibinfo  {journal} {JPS Conf. Proc.}\ }\textbf {\bibinfo {volume} {18}},\
  \bibinfo {pages} {011002} (\bibinfo {year} {2017})},\ \Eprint
  {http://arxiv.org/abs/1708.03213} {arXiv:1708.03213 [hep-ph]} \BibitemShut
  {NoStop}%
%%CITATION = ARXIV:1708.03213;%%
\bibitem [{\citenamefont {Will}(1993)}]{will1993}%
  \BibitemOpen
  \bibfield  {author} {\bibinfo {author} {\bibfnamefont {C.~M.}\ \bibnamefont
  {Will}},\ }\href@noop {} {\emph {\bibinfo {title} {{Theory and experiment in
  gravitational physics}}}}\ (\bibinfo  {publisher} {Cambridge, University
  Press, Cambridge, England},\ \bibinfo {year} {1993})\BibitemShut {NoStop}%
%%CITATION = INSPIRE-357130;%%
\bibitem [{\citenamefont {Kosteleck\'y}\ and\ \citenamefont
  {Tasson}(2009)}]{akjt2009}%
  \BibitemOpen
  \bibfield  {author} {\bibinfo {author} {\bibfnamefont {V.~A.}\ \bibnamefont
  {Kosteleck\'y}}\ and\ \bibinfo {author} {\bibfnamefont {J.}~\bibnamefont
  {Tasson}},\ }\href {\doibase 10.1103/PhysRevLett.102.010402} {\bibfield
  {journal} {\bibinfo  {journal} {Phys. Rev. Lett.}\ }\textbf {\bibinfo
  {volume} {102}},\ \bibinfo {pages} {010402} (\bibinfo {year} {2009})},\
  \Eprint {http://arxiv.org/abs/0810.1459} {arXiv:0810.1459 [gr-qc]}
  \BibitemShut {NoStop}%
%%CITATION = ARXIV:0810.1459;%%
\bibitem [{\citenamefont {Shen}\ \emph {et~al.}(2019)\citenamefont {Shen},
  \citenamefont {Yang}, \citenamefont {Guo},\ and\ \citenamefont
  {Zhang}}]{iop}%
  \BibitemOpen
  \bibfield  {author} {\bibinfo {author} {\bibfnamefont {W.}~\bibnamefont
  {Shen}}, \bibinfo {author} {\bibfnamefont {Z.}~\bibnamefont {Yang}}, \bibinfo
  {author} {\bibfnamefont {Z.}~\bibnamefont {Guo}}, \ and\ \bibinfo {author}
  {\bibfnamefont {W.}~\bibnamefont {Zhang}},\ }\href {\doibase
  10.1016/j.geog.2019.03.001} {\bibfield  {journal} {\bibinfo  {journal}
  {Geod.\ Geodyn.}\ }\textbf {\bibinfo {volume} {10}},\ \bibinfo {pages} {118}
  (\bibinfo {year} {2019})}\BibitemShut {NoStop}%
\bibitem [{\citenamefont {Riehle}\ \emph {et~al.}(1991)\citenamefont {Riehle},
  \citenamefont {Kisters}, \citenamefont {Witte}, \citenamefont {Helmcke},\
  and\ \citenamefont {Bord\'e}}]{riehle91}%
  \BibitemOpen
  \bibfield  {author} {\bibinfo {author} {\bibfnamefont {F.}~\bibnamefont
  {Riehle}}, \bibinfo {author} {\bibfnamefont {T.}~\bibnamefont {Kisters}},
  \bibinfo {author} {\bibfnamefont {A.}~\bibnamefont {Witte}}, \bibinfo
  {author} {\bibfnamefont {J.}~\bibnamefont {Helmcke}}, \ and\ \bibinfo
  {author} {\bibfnamefont {C.~J.}\ \bibnamefont {Bord\'e}},\ }\href {\doibase
  10.1103/PhysRevLett.67.177} {\bibfield  {journal} {\bibinfo  {journal} {Phys.
  Rev. Lett.}\ }\textbf {\bibinfo {volume} {67}},\ \bibinfo {pages} {177}
  (\bibinfo {year} {1991})}\BibitemShut {NoStop}%
\bibitem [{\citenamefont {Gustavson}\ \emph {et~al.}(1997)\citenamefont
  {Gustavson}, \citenamefont {Bouyer},\ and\ \citenamefont
  {Kasevich}}]{gustavson97}%
  \BibitemOpen
  \bibfield  {author} {\bibinfo {author} {\bibfnamefont {T.~L.}\ \bibnamefont
  {Gustavson}}, \bibinfo {author} {\bibfnamefont {P.}~\bibnamefont {Bouyer}}, \
  and\ \bibinfo {author} {\bibfnamefont {M.~A.}\ \bibnamefont {Kasevich}},\
  }\href {\doibase 10.1103/PhysRevLett.78.2046} {\bibfield  {journal} {\bibinfo
   {journal} {Phys. Rev. Lett.}\ }\textbf {\bibinfo {volume} {78}},\ \bibinfo
  {pages} {2046} (\bibinfo {year} {1997})}\BibitemShut {NoStop}%
\bibitem [{\citenamefont {Stockton}\ \emph {et~al.}(2011)\citenamefont
  {Stockton}, \citenamefont {Takase},\ and\ \citenamefont
  {Kasevich}}]{stockton11}%
  \BibitemOpen
  \bibfield  {author} {\bibinfo {author} {\bibfnamefont {J.~K.}\ \bibnamefont
  {Stockton}}, \bibinfo {author} {\bibfnamefont {K.}~\bibnamefont {Takase}}, \
  and\ \bibinfo {author} {\bibfnamefont {M.~A.}\ \bibnamefont {Kasevich}},\
  }\href {\doibase 10.1103/PhysRevLett.107.133001} {\bibfield  {journal}
  {\bibinfo  {journal} {Phys. Rev. Lett.}\ }\textbf {\bibinfo {volume} {107}},\
  \bibinfo {pages} {133001} (\bibinfo {year} {2011})}\BibitemShut {NoStop}%
\bibitem [{\citenamefont {Savoie}\ \emph {et~al.}(2018)\citenamefont {Savoie},
  \citenamefont {Altorio}, \citenamefont {Fang}, \citenamefont {Sidorenkov},
  \citenamefont {Geiger},\ and\ \citenamefont {Landragin}}]{savoie18}%
  \BibitemOpen
  \bibfield  {author} {\bibinfo {author} {\bibfnamefont {D.}~\bibnamefont
  {Savoie}}, \bibinfo {author} {\bibfnamefont {M.}~\bibnamefont {Altorio}},
  \bibinfo {author} {\bibfnamefont {B.}~\bibnamefont {Fang}}, \bibinfo {author}
  {\bibfnamefont {L.~A.}\ \bibnamefont {Sidorenkov}}, \bibinfo {author}
  {\bibfnamefont {R.}~\bibnamefont {Geiger}}, \ and\ \bibinfo {author}
  {\bibfnamefont {A.}~\bibnamefont {Landragin}},\ }\href {\doibase
  10.1126/sciadv.aau7948} {\bibfield  {journal} {\bibinfo  {journal} {Sci.\
  Adv.}\ }\textbf {\bibinfo {volume} {4}},\ \bibinfo {pages} {eaau7948}
  (\bibinfo {year} {2018})}\BibitemShut {NoStop}%
\bibitem [{usn()}]{usnavy}%
  \BibitemOpen
  \href@noop {} {}\bibinfo {note} {{United States Naval Observatory}, web-site,
  \url{https://aa.usno.navy.mil/data/docs/EarthSeasons.php}, Accessed: May 2,
  2019}\BibitemShut {NoStop}%
\bibitem [{\citenamefont {Ding}\ and\ \citenamefont
  {Kosteleck\'y}(2016)}]{akyd}%
  \BibitemOpen
  \bibfield  {author} {\bibinfo {author} {\bibfnamefont {Y.}~\bibnamefont
  {Ding}}\ and\ \bibinfo {author} {\bibfnamefont {V.~A.}\ \bibnamefont
  {Kosteleck\'y}},\ }\href {\doibase 10.1103/PhysRevD.94.056008} {\bibfield
  {journal} {\bibinfo  {journal} {Phys. Rev. D}\ }\textbf {\bibinfo {volume}
  {94}},\ \bibinfo {pages} {056008} (\bibinfo {year} {2016})},\ \Eprint
  {http://arxiv.org/abs/1608.07868} {arXiv:1608.07868 [hep-ph]} \BibitemShut
  {NoStop}%
%%CITATION = ARXIV:1608.07868;%%
\bibitem [{\citenamefont {Di~Virgilio}\ \emph {et~al.}(2017)\citenamefont
  {Di~Virgilio}, \citenamefont {Belfi}, \citenamefont {Ni}, \citenamefont
  {Beverini}, \citenamefont {Carelli}, \citenamefont {Maccioni},\ and\
  \citenamefont {Porzio}}]{diVirgilio2017}%
  \BibitemOpen
  \bibfield  {author} {\bibinfo {author} {\bibfnamefont {A.~D.~V.}\
  \bibnamefont {Di~Virgilio}}, \bibinfo {author} {\bibfnamefont
  {J.}~\bibnamefont {Belfi}}, \bibinfo {author} {\bibfnamefont {W.-T.}\
  \bibnamefont {Ni}}, \bibinfo {author} {\bibfnamefont {N.}~\bibnamefont
  {Beverini}}, \bibinfo {author} {\bibfnamefont {G.}~\bibnamefont {Carelli}},
  \bibinfo {author} {\bibfnamefont {E.}~\bibnamefont {Maccioni}}, \ and\
  \bibinfo {author} {\bibfnamefont {A.}~\bibnamefont {Porzio}},\ }\href
  {\doibase 10.1140/epjp/i2017-11452-6} {\bibfield  {journal} {\bibinfo
  {journal} {Eur. Phys. J. Plus}\ }\textbf {\bibinfo {volume} {132}},\ \bibinfo
  {pages} {157} (\bibinfo {year} {2017})}\BibitemShut {NoStop}%
\bibitem [{\citenamefont {Chung}\ \emph {et~al.}(2009)\citenamefont {Chung},
  \citenamefont {Chiow}, \citenamefont {Herrmann}, \citenamefont {Chu},\ and\
  \citenamefont {Muller}}]{chung2009}%
  \BibitemOpen
  \bibfield  {author} {\bibinfo {author} {\bibfnamefont {K.-Y.}\ \bibnamefont
  {Chung}}, \bibinfo {author} {\bibfnamefont {S.-W.}\ \bibnamefont {Chiow}},
  \bibinfo {author} {\bibfnamefont {S.}~\bibnamefont {Herrmann}}, \bibinfo
  {author} {\bibfnamefont {S.}~\bibnamefont {Chu}}, \ and\ \bibinfo {author}
  {\bibfnamefont {H.}~\bibnamefont {Muller}},\ }\href {\doibase
  10.1103/PhysRevD.80.016002} {\bibfield  {journal} {\bibinfo  {journal} {Phys.
  Rev. D}\ }\textbf {\bibinfo {volume} {80}},\ \bibinfo {pages} {016002}
  (\bibinfo {year} {2009})},\ \Eprint {http://arxiv.org/abs/0905.1929}
  {arXiv:0905.1929 [gr-qc]} \BibitemShut {NoStop}%
%%CITATION = ARXIV:0905.1929;%%
\bibitem [{\citenamefont {Muller}\ \emph {et~al.}(2008)\citenamefont {Muller},
  \citenamefont {Chiow}, \citenamefont {Herrmann}, \citenamefont {Chu},\ and\
  \citenamefont {Chung}}]{muller2007}%
  \BibitemOpen
  \bibfield  {author} {\bibinfo {author} {\bibfnamefont {H.}~\bibnamefont
  {Muller}}, \bibinfo {author} {\bibfnamefont {S.-w.}\ \bibnamefont {Chiow}},
  \bibinfo {author} {\bibfnamefont {S.}~\bibnamefont {Herrmann}}, \bibinfo
  {author} {\bibfnamefont {S.}~\bibnamefont {Chu}}, \ and\ \bibinfo {author}
  {\bibfnamefont {K.-Y.}\ \bibnamefont {Chung}},\ }\href {\doibase
  10.1103/PhysRevLett.100.031101} {\bibfield  {journal} {\bibinfo  {journal}
  {Phys. Rev. Lett.}\ }\textbf {\bibinfo {volume} {100}},\ \bibinfo {pages}
  {031101} (\bibinfo {year} {2008})},\ \Eprint {http://arxiv.org/abs/0710.3768}
  {arXiv:0710.3768 [gr-qc]} \BibitemShut {NoStop}%
%%CITATION = ARXIV:0710.3768;%%
\bibitem [{\citenamefont {{Kosteleck{\'y}}}\ and\ \citenamefont
  {{Tasson}}(2015)}]{akjt15}%
  \BibitemOpen
  \bibfield  {author} {\bibinfo {author} {\bibfnamefont {V.~A.}\ \bibnamefont
  {{Kosteleck{\'y}}}}\ and\ \bibinfo {author} {\bibfnamefont {J.~D.}\
  \bibnamefont {{Tasson}}},\ }\href {\doibase 10.1016/j.physletb.2015.08.060}
  {\bibfield  {journal} {\bibinfo  {journal} {Phys. Lett. B}\ }\textbf
  {\bibinfo {volume} {749}},\ \bibinfo {pages} {551} (\bibinfo {year}
  {2015})},\ \Eprint {http://arxiv.org/abs/1508.07007} {arXiv:1508.07007
  [gr-qc]} \BibitemShut {NoStop}%
\end{thebibliography}%

\end{document}